\documentclass[pre,aps,preprint,showpacs]{revtex4-1}
\usepackage{revsymb4-1,latexsym,amssymb,amsmath,comment}
\usepackage{graphicx,psfrag,subfigure,stmaryrd}
\usepackage[ansinew]{inputenc}

\newcommand{\be}{\begin{equation}}
\newcommand{\ee}{\end{equation}}
\newcommand{\bel}[1]{\begin{equation}\label{#1}}
\newcommand{\bea}{\begin{eqnarray}}
\newcommand{\eea}{\end{eqnarray}}
\newcommand{\ba}{\begin{array}}
\newcommand{\ea}{\end{array}}
\newcommand{\bef}{\begin{figure}}
\newcommand{\ef}{\end{figure}}

\begin{document}

\author{Thomas Bose and Steffen Trimper}
\affiliation{Institute of Physics,
Martin-Luther-University, D-06099 Halle, Germany}
\email{thomas.bose@physik.uni-halle.de}
\email{steffen.trimper@physik.uni-halle.de}
\title{Correlation Effects in Stochastic Ferromagnetic Systems}
\date{\today }

\begin{abstract}

We analyze the Landau-Lifshitz-Gilbert equation when the precession motion of the magnetic moments is additionally subjected to an uniaxial anisotropy 
and is driven by a multiplicative coupled stochastic field with a finite correlation time $\tau$. The mean value for the spin wave 
components offers that the spin-wave dispersion relation and its damping is strongly influenced by the deterministic Gilbert damping parameter $\alpha$, the 
strength of the stochastic forces $D$ and its temporal range $\tau$. The spin-spin-correlation function can be calculated in the low correlation time limit by
deriving an evolution equation for the joint probability function. The stability analysis enables us to find the phase diagram within the $\alpha-D$ plane for 
different values of $\tau$ where damped spin wave solutions are stable. Even for zero deterministic Gilbert damping the magnons offer a finite lifetime. We detect 
a parameter range where the deterministic and the stochastic damping mechanism are able to compensate each other leading to undamped spin-waves. The onset is
characterized by a critical value of the correlation time. An enhancement of $\tau$ leads to an increase of the oscillations of the correlation function.

\end{abstract}

\maketitle

\section{Introduction}

\noindent Magnetism can be generally characterized and analyzed on different length and time scales. The description of fluctuations of the
magnetization, the occurrence of damped spin waves and the influence of additional stochastic forces are successfully performed on a
mesoscopic scale where the spin variables are represented by a continuous spatio-temporal variable  
\cite{Landau:ElecContMed:Book:1989}. In this case 
a well established approach is based upon the Landau-Lifshitz equation \cite{Landau:ZdS:8:p153:1935} which describes the precession motion
of the magnetization in 
an effective magnetic field. This field consists of a superposition of an external field and 
internal fields, produced by the interacting magnetic moments. The latter one is strongly influenced by the isotropic exchange interaction
and the magnetocrystalline anisotropy, for a recent review see \cite{Tserkovnyak:RoMP:77:p1375:2005}. The studies using this frame are
concentrated on different dynamical aspects as the switching behavior of magnetic nanoparticles which can be controlled by external
time-dependent magnetic fields \cite{Sukhov:JoPM:20:p125226:2008} and spin-polarized electric currents  
\cite{Slonczewski:JoMaMM:159:p1:1996, Berger:PRB:54:p9353:1996}. Such a current-induced spin transfer allows the manipulation of magnetic
nanodevices. Recently, it has been demonstrated 
that an electric current, flowing through a magnetic bilayer, can induce a coupling between the 
layers \cite{Urazhdin:PRB:78:p60405:2008}. Likewise, such a current 
can also cause the motion of magnetic domain walls in a nanowire 
\cite{Kruger:PRB:75:p54421:2007}. 
Another aspect is the dynamical response of ferromagnetic nanoparticles as probed by ferromagnetic resonance, studied in 
\cite{Usadel:PRB:73:p212405:2006}. In describing all this more complex behavior of magnetic systems, the Landau-Lifshitz equation 
has to be extended by the inclusion of dissipative processes. A damping term is introduced phenomenologically in such a manner, that the
magnitude of the magnetization $\vec S$ is preserved at any time. Furthermore, 
the magnetization should align with the effective field in the long time limit. A realization is given by \cite{Landau:ZdS:8:p153:1935} 
\be 
\frac{\partial \bf{S}}{\partial t} = - \gamma [\, \bf{S} \times \bf{B}_{\textrm{eff}} \,] -  
\varepsilon   \left[\, \bf{S} \times ( \bf{
S} \times \bf{B}_{\textrm{eff}})\, \right]\,.
\label{llg}  
\ee
The quantities $\gamma$ and $\varepsilon $ are the gyromagnetic ratio and the damping parameter, respectively. 
An alternative equation for the magnetization dynamics had been proposed by Gilbert \cite{ Gilbert:ITOM:40:p3443:2004}. The Gilbert
equation yields an implicit form of the evolution of the magnetization. A combination of both equations, called Landau-Lifshitz-Gilbert
equation (LLG) will be used as the basic relation for our studies, see Eq.~\eqref{LLG1}. The origin of the damping term as a
non-relativistic expansion of the Dirac equation has been discussed in 
\cite{Hickey:PRL:102:p137601:2009} and a generalization of the LLG for 
conducting ferromagnetics is offered in \cite{Zhang:PRL:102:p86601:2009}. The form of the damping seems to be quite general as it has been
demonstrated in \cite{Trimper:PRB:76:p94108:2007} using symmetry arguments for ferroelectric systems.\\
\noindent As a new aspect let us focus on the influence of stochastic fields. The interplay between current and magnetic fluctuations and
dissipation has been studied recently in \cite{Foros:PRB:79:p214407:2009}. Via the spin-transfer torque, spin-current noise causes a 
significant enhancement of the magnetization fluctuations. Such a spin polarized current may transfer momentum to a magnet which leads 
to a spin-torque phenomenon. The shot noise associated with the 
current gives rise to a stochastic force \cite{Chudnovskiy:PRL:101:p66601:2008}. In our paper we discuss the interplay between different 
dissipation mechanism, namely the inherent deterministic damping in Eq.~\eqref{llg} and the stochastic magnetic field originated for instance by defect configurations giving rise to a different coupling strength between the magnetic moments. Assuming further, that the stochastic magnetic field is characterized by a finite correlation time, the system offers memory effects which might lead to a decoherent spin precession. To that aim we analyze a ferromagnet in the classical limit, i.e., the magnetic order is referred to single magnetic atoms which occupy equivalent crystal positions, and the mean values of their spins exhibit a parallel orientation. The last one is caused by the isotropic exchange interaction which will be here supplemented by a magneto-crystalline anisotropy that defines the direction of the preferred orientation. Especially, we discuss the influence 
of an uniaxial anisotropy. The coupling between different dissipation mechanisms, mentioned above, leads to pronounced correlations, which are discussed below. Due to the multiplicative coupling of 
the stochastic field and the finite correlation time the calculation of the spin-spin correlation function is more complicated. To that aim we have to derive an equivalent evolution equation for the joint probability distribution function. Within the small correlation time limit this approach can be fulfilled in an analytical manner. Our analysis is related to a recent paper \cite{Basko:PRB:79:p64418:2009} in which likewise the stochastic dynamics of the magnetization in ferromagnetic nanoparticles has been studied. Further, we refer also to a recent paper \cite{Denisov:PRB:75:p184432:2007} where the mean first passage time and the relaxation of magnetic moments has been analyzed. Different to those papers our approach is concentrated on the correlation effects in stochastic system with colored noise.\\
\noindent Our paper is organized as follows: In Sec.II we discuss the LLG and characterize the additional stochastic field. The equations for the single and the two particle joint probability distribution are derived in Sec.III. Using these functions we obtain the mean value of the spin wave variable and the spin-spin correlation function. The phase diagram, based on the stability analysis, is presented in Sec.IV. In Sec.V we finish with some conclusions.

\section{Model}

\noindent In order to develop a stochastic model for the spin dynamics in ferromagnetic systems let us first consider the deterministic part of the equation of motion. We focus on a description 
based upon the level of Landau-Lifshitz phenomenology \cite{Landau:ZdS:8:p153:1935}, for a recent review see \cite{Tserkovnyak:RoMP:77:p1375:2005}. To follow this line we consider a high spin  systems in a ferromagnet sufficiently below the Curie temperature. In that regime the dynamics of the magnet are dominated by transverse fluctuations of the spatio-temporal varying local  magnetization. The weak excitations, called spin waves or magnons, are determined by a dispersion relation, the wavelength of which should be large compared to the lattice constant $a$, i.e., the relation \mbox{$q\cdot a\ll 1$} is presumed to be satisfied, where $q$ is the wavenumber. 
In this limit the direction of the spin varies slowly while its magnitude
\mbox{$|\mathbf{S}|=m_s$} remains constant in time. A proper description for such a situation is achieved by applying the Landau-Lifshitz-Gilbert equation (LLG) \cite{Gilbert:ITOM:40:p3443:2004,Daniel:PAMaIA:120:p125:1983,Sukhov:JoPM:20:p125226:2008}. 
The spin variable is represented by $\mathbf{S} = m_s  \mathbf{\hat{n}}$, where $\mathbf{\hat{n}}(\mathbf{r},t)$ is 
a continuous variable which characterizes the local orientation of the magnetic moment. The evolution equation for that local orientation reads
\be
\frac{\partial \mathbf{\hat{n}}}{\partial t}=-\frac{\gamma}{1+\alpha ^2}\,\mathbf{\hat{n}}\times \left[ \mathbf{B_{\textrm{eff}}}+\alpha \,[\bf{\hat{n}}\times \mathbf{B_{\textrm{eff}}}]\right] \,.
\label{LLG1}
\ee
The quantities $\gamma$ and $\alpha$ are the gyromagnetic ratio and the dimensionless Gilbert damping parameter, respectively, where $\alpha$ is related to $\varepsilon $ introduced in 
Eq.~\eqref{llg}. $\mathbf{B_{\textrm{eff}}}$ is the effective magnetic field that drives the motion of the spin density. Generally, it consists of an internal part originated by the interaction of the spins and an external field. This effective field is related to the Hamiltonian of the system 
by functional variation with respect to $\mathbf{\hat{n}}$
\be
\mathbf{B_{\textrm{eff}}} = - m_s^{-1} \frac{\delta \mathcal{H}}{\delta \mathbf{\hat n}}\,.  
\label{eff}
\ee
In absence of an external field the Hamiltonian can be expressed as  \cite{Lakshmanan:PRL:53:p2497:1984,Bar'Yakhtar:DynTopMagSol:Book:1994}
\be
\begin{aligned}
\mathcal{H}=\int{d^3\mathbf{r}\,\{ w_{ex}+w_{an}\}}\,, \qquad \textrm{with} \\
w_{ex}=\frac{1}{2}\,m_s\,\kappa \,(\nabla \mathbf{\hat{n}})^2 \quad \textrm{and} \quad w_{an}=\frac{1}{2}\,m_s\,\Gamma \,\sin ^2\,\theta \,.
\end{aligned}
\label{fieldhamil}
\ee
Thereby, the constants $\kappa$ and $\Gamma$ denote the exchange energy density and the magneto-crystalline anisotropy energy density. 
To be more precise, \mbox{$\kappa \propto Ja^2$}, $J$ being the coupling strength that measures the interaction between nearest neighbors in the isotropic Heisenberg model \cite{Lakshmanan:PA:84:p577:1976}. Once again $a$ is the lattice constant. Notice that the form of the  exchange energy in the Hamiltonian~\eqref{fieldhamil} arises from the Heisenberg model in the classical limit. 
The quantity $\theta$ represents the angle between $\mathbf{\hat{n}}$ and the anisotropy axis \mbox{$\boldsymbol{\hat{\nu}}=(0,0,1)$}, where $\boldsymbol{\hat{\nu}}$ points in the direction of the easy axis in the ground state in the case of zero applied external field. Thus, the constant $\Gamma >0$ characterizes anisotropy as a consequence of relativistic interactions (spin-orbital and dipole-dipole ones \cite{Bar'Yakhtar:DynTopMagSol:Book:1994}). In deriving Eq.~\eqref{fieldhamil} we have used $\mathbf{\hat{n}}^2 = 1$. Although it is more conventional to introduce the angular coordinates $(\theta,\Phi)$ \cite{Landau:ZdS:8:p153:1935,Sukhov:JoPM:20:p125226:2008}, 
we find it more appropriate to use Cartesian coordinates. To proceed, we divide the vector $\mathbf{\hat{n}}$ into a static and a dynamic part designated by $\boldsymbol{\mu}$ and $\boldsymbol{\varphi}$, respectively. In the linearized spin wave approach let us make the ansatz
\be
\mbox{$\mathbf{\hat{n}}(\mathbf{r},t)$} =\boldsymbol{\mu}(\mathbf{r}) + \boldsymbol{\varphi} (\mathbf{r},t) = \mu \,\boldsymbol{\hat{\nu}}+\boldsymbol{\varphi} \,,\quad \mu = \textrm{const}.\,,
\label{ansatz}
\ee 
where $\mathbf{\hat{n}}^2=1$ is still valid. The effective field can now be obtained from Eqs.~\eqref{eff} and \eqref{fieldhamil}. This yields
\be
\mathbf{B_{\textrm{eff}}}=\kappa\,\nabla ^2\,\boldsymbol{\varphi}-\Gamma \,\boldsymbol{\varphi '}; 
\quad\quad \boldsymbol{\varphi '} = (\varphi _{1},\varphi _{2},0) \,.
\label{Beff1}
\ee
Eq.~\eqref{LLG1} together with Eqs.~\eqref{eff} and \eqref{fieldhamil} represent the deterministic model for a classical ferromagnet. 
In order to extent the model let us supplement the effective magnetic field in Eq.~\eqref{Beff1} by a stochastic component yielding an effective random field \mbox{$\mathbf{B_{\textrm{eff}}}=\mathbf{B_{\textrm{eff}}}+\boldsymbol{\eta}(t)$}. The stochastic process $\boldsymbol{\eta}(t)$ is assumed to be Gaussian distributed with zero mean and obeying a colored correlation function
\be
\tilde{\chi}_{ij}(t,t')=\langle \eta _i(t)\,\eta _j(t')\rangle =\frac{\tilde{D}_{ij}}{\tilde{\tau}_{ij}}\,\exp\left[-\frac{\mid t-t'\mid}{\tilde{\tau}_{ij}}\right] \,.
\label{noise1}
\ee 
Here, $\tilde{D}_{ij}$ and $\tilde{\tau}_{ij}$ are the noise strength and the finite correlation time of the noise $\boldsymbol{\eta}$. Due to the coupling of the effective field to the spin orientation $\mathbf{\hat{n}}$ the stochastic process is a multiplicative one. 
Microscopically, such a random process might be originated by a fluctuating coupling strength for instance. The situation associated with 
our model is illustrated in Fig.~\ref{Ferro1} and can be understood as follows: 
The stochastic vector field $\boldsymbol{\eta}(t)$ is able to change the orientation of the localized moment at different times. Therefore, fixed phase relations between adjacent spins might be destroyed. Moreover, the $\boldsymbol{\eta}(t_k)$ are interrelated due to the finite correlation time $\tau$. The anisotropy axis defines the preferred orientation of the mean value of magnetization. 
\bef
	\includegraphics[width=10cm]{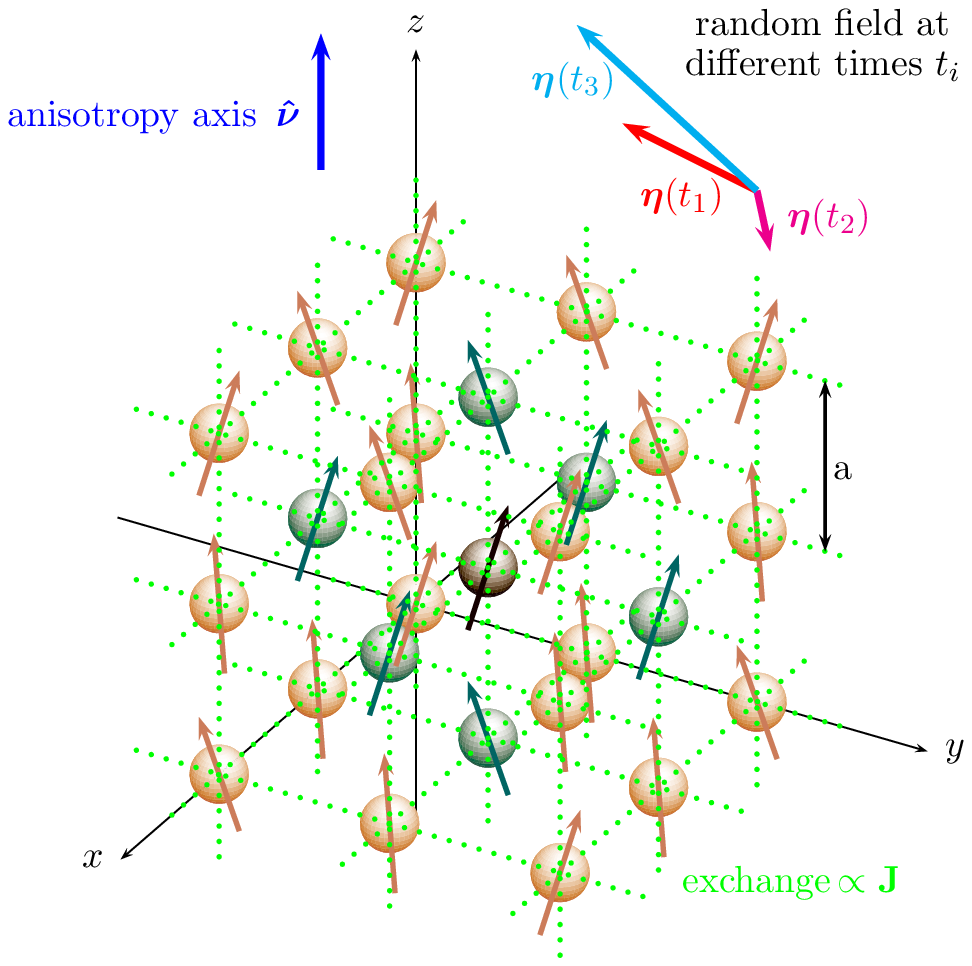}
	\caption{Part of a ferromagnetic domain influenced by stochastic forces for the example of cubic symmetry with lattice constant $a$. The black spin in the center only interacts with its nearest neighbors (green), where $J$ is a measure for the exchange integral.} 
\label{Ferro1}
\ef
Due to the inclusion of $\boldsymbol{\eta}(t)$ the deterministic Eq.~\eqref{LLG1} is transformed into the stochastic LLG. Using Eq.~\eqref{ansatz} it follows
\be
\frac{\partial \boldsymbol{\varphi}}{\partial t}=-\frac{\gamma}{1+\alpha ^2}\,(\boldsymbol{\mu}+\boldsymbol{\varphi})\times \left[ \mathbf{B_{\textrm{eff}}}+\alpha \,[(\boldsymbol{\mu}+\boldsymbol{\varphi})\times \mathbf{B_{\textrm{eff}}}]\right] \,.
\label{LLG2}
\ee
The random magnetic field is defined by
\be
\mathbf{B_{\textrm{eff}}}=\kappa\,\nabla ^2\,\boldsymbol{\varphi}-\Gamma \,\boldsymbol{\varphi '} 
+ \boldsymbol{\eta}(t) \,,
\label{beff2}
\ee
where $\boldsymbol{\varphi '}$ is given in Eq.~\eqref{Beff1}. With regard to the following procedure we suppose the random field to be solely generated dynamically, i.e., \mbox{$\mathbf{\hat{n}}\times \boldsymbol{\eta}(t)=\boldsymbol{\varphi}\times \boldsymbol{\eta}(t)$}. So far, the dynamics of our model (Eqs.~\eqref{LLG2} and \eqref{beff2}) are reflected by a nonlinear, stochastic partial differential equation (PDE). Using Fourier transformation, 
i.e., \mbox{$\boldsymbol{\psi}(\mathbf{q},t)=\mathcal{F}\{ \boldsymbol{\varphi}(\mathbf{r},t)\}$} and introducing the following dimensionless quantities 
\be
\beta =(l_0\,q)^2+1 \quad ,\quad l_0^2=\frac{\kappa }{\Gamma} \quad ,\quad \omega =\gamma \,\Gamma \quad ,\quad \bar{t}=\omega \,t \quad ,\quad \boldsymbol\lambda (t)=\frac{\boldsymbol\eta (t)}{\Gamma }\,,
\label{rel1}
\ee
the components $\psi _i(\mathbf{q},t)$ fulfill the equation 
\be
\frac{d}{dt}\psi _i(\mathbf{q},t)=\Omega _i(\boldsymbol{\psi}(\mathbf{q},t))+\Lambda _{ij}(\boldsymbol{\psi}(\mathbf{q},t))\,\boldsymbol{\lambda}_j(t)\,.
\label{sys}
\ee
The quantity $l_0$ is the characteristic magnetic length \cite{Kosevich:PRSoPL:194:p117:1990}. The vector $\boldsymbol{\Omega}$ and the matrix $\Lambda $ are given by
\be
\boldsymbol{\Omega}=\xi \,\mu \,\beta \begin{pmatrix}
										-(\alpha \mu \,\psi _1+ \psi _2) \\
											\psi _1-\alpha \mu \,\psi _2 \\
											0
									\end{pmatrix}
\,,\qquad \xi =\frac{1}{1+\alpha ^2}\,,
\label{omega}
\ee
and
\be
\Lambda =\xi \begin{pmatrix}
						\alpha \mu \,\psi _3 & \psi _3 & -(\psi _2+\alpha \mu \,\psi _1) \\
						-\psi _3 & \alpha \mu \,\psi _3 & \psi _1-\alpha \mu \,\psi _2 \\
						\psi _2 & -\psi _1 & 0 
					\end{pmatrix}\,.
\label{Lambda}
\ee
For convenience we have substituted \mbox{$\bar{t}\rightarrow t$} again. The statistical properties of $\boldsymbol{\lambda}(t)$ are expressed as \mbox{$\langle \boldsymbol{\lambda}(t)\rangle=0$} and
\be
\chi _{kl}(t,t')=\langle \lambda _k(t)\,\lambda _l(t')\rangle =\frac{D_{kl}}{\tau _{kl}}\,\delta _{kl}\,\exp\left[-\frac{\mid t-t'\mid}{\tau _{kl}}\right] \,\xrightarrow[]{\tau _{kl}\rightarrow 0} 2\,D_{kl}\,\delta _{kl}\,\delta (t-t') \,.
\label{noise2}
\ee
Incidentally, in the limit \mbox{$\tau \rightarrow 0$} the usual white noise properties are recovered. We emphasize that although we regard the long-wavelength limit \mbox{($a\cdot q\ll 1$)}, wave vectors for which \mbox{$l_0\cdot q\gg 1$} (in Eq.~\eqref{rel1}) can also occur \cite{Kosevich:PRSoPL:194:p117:1990}. But this case is not discussed in the present paper and will be the content of future work. Whereas, in what follows we restrict our considerations to the case \mbox{$q\rightarrow 0$} so that, actually, \mbox{$l_0\cdot q\ll 1$} is fulfilled. Hence, we can set \mbox{$\beta =1$} approximately in Eq.~\eqref{rel1}. Due to the anisotropy the spin wave dispersion relation offers a gap at $\mathbf{q} = 0$. Owing to this fact $\boldsymbol{\psi}$ is studied at zero wave vector. For this situation the assumption 
of a space-independent stochastic force $\eta_i(t)$, compare Eq.~\eqref{noise1}, is reasonable. For non-zero wave vector the 
noise field should be a spatiotemporal field $\eta_i((\mathbf{r},t)$. Because our model is based on a short range interaction we expect that the corresponding noise correlation function is $\delta$-correlated, i.e. instead of \eqref{noise2} we have 
$$
\chi _{kl}(\mathbf{r},t; \mathbf{r'}, t') = \frac{D_{kl}}{\tau _{kl}}\,\delta _{kl}\,\exp\left[-\frac{\mid t-t'\mid}{\tau _{kl}}\right] 
2 M \delta (\mathbf{r} - \mathbf{r'})\,,
$$
where $M$ is the strength of the spatial correlation. Using this relation we are able to study also the case of small $\mathbf{q}$ which 
satisfies \mbox{$l_0\cdot q\ll 1$}. In the present paper we concentrate on the case of zero wave vector $\mathbf{q} = 0$.

\section{Correlation functions}

\noindent In the present section let us discuss the statistical behavior of the basic Eqs.~\eqref{sys}-\eqref{noise2}. They describe a  non-stationary, non-Markovian process attributed to the finite correlation time. Due to their common origin both characteristics 
can not be analyzed separately. In the limit $\tau \rightarrow 0$, Eq.~\eqref{sys} defines a Markovian process which provides also 
stationarity by an appropriate choice of initial conditions \cite{Hernandez-Machado:JoMP:25:p1066:1984}. However, the present 
study is focused on the effect of nonzero correlation times. To that purpose we need a proper probability distribution function 
which reflects the stochastic process defined by Eqs.~\eqref{sys}-\eqref{noise2}. In deriving the relevant joint probability distribution 
function we follow the line given in \cite{Hernandez-Machado:ZFPBM:52:p335:1983}, where the detailed calculations had been carried out, see also the references cited therein. In particular, it has been underlined in those papers that in order to calculate correlation functions of type $\langle \psi _i(t)\,\psi _j(t')\rangle$ a single probability distribution function $P(\boldsymbol{\psi},t)$ is not sufficient. 
Instead of that one needs a joint probability distribution of the form $P(\boldsymbol{\psi},t;\boldsymbol{\psi '},t')$. Before proceeding 
let us shortly summarize the main steps to get the joint probability distribution function. To simplify the calculation we assume $\tau _{kl}=\tau \,\delta _{kl}$ and $D_{kl}=D\,\delta _{kl}$. Notice that our system has no ergodic properties what would directly allow us to relate the stochastic interferences with temperature fluctuations by means of a fluctuation-dissipation theorem. Based on 
Eq.~\eqref{sys} the appropriate joint probability distribution is defined by \cite{Hernandez-Machado:ZFPBM:52:p335:1983,Kampen:BJoP:28:p90:1998}, for a more general discussion compare also 
\cite{Kampen:StochProcPhysChem:Book:1981}:
\be
P(\boldsymbol{\psi},t;\boldsymbol{\psi '},t')=\left\langle \delta (\boldsymbol{\psi}(t)-\boldsymbol{\psi})\,\delta (\boldsymbol{\psi}(t')-\boldsymbol{\psi'})\right\rangle \,.
\label{prob1}
\ee
Here the average is performed over all realizations of the stochastic process. In defining the joint probability distribution 
function we follow the convention to indicate the stochastic process by the function $\boldsymbol{\psi}(t)$ whereas the quantity 
without arguments $\boldsymbol{\psi}$ stands for the special values of the stochastic variable. These values are even relalized with 
the probaility $P(\boldsymbol{\psi},t;\boldsymbol{\psi '},t')$. The equation of motion for this probability distribution reads 
according to \cite{Hernandez-Machado:ZFPBM:52:p335:1983}
\begin{align}
\begin{aligned}
\frac{\partial}{\partial t}&P(\boldsymbol{\psi},t;\boldsymbol{\psi '},t')\\
						 = &-\frac{\partial}{\partial \psi _i}\int\limits_{0}^{t} \chi _{jk}(t,t_1)\,\left\langle \left[ \frac{\delta\,\psi _i(t)}{\delta\,\lambda _k(t_1)}\right] _{\boldsymbol{\psi}(t)=\boldsymbol{\psi}}\cdot \delta (\boldsymbol{\psi}(t)-\boldsymbol{\psi})\,\delta (\boldsymbol{\psi}(t')-\boldsymbol{\psi'})\vphantom{\frac{\delta \psi _i}{\delta \lambda _k}}\right\rangle \,dt_1\\
						   &-\frac{\partial}{\partial {\psi '}_i}\int\limits_{0}^{t'} \chi _{jk}(t,t_1)\,\left\langle \left[ \frac{\delta\,\psi _i(t')}{\delta\,\lambda _k(t_1)}\right] _{\boldsymbol{\psi}(t')=\boldsymbol{\psi '}}\cdot \delta (\boldsymbol{\psi}(t)-\boldsymbol{\psi})\,\delta (\boldsymbol{\psi}(t')-\boldsymbol{\psi '})\vphantom{\frac{\delta \psi _i}{\delta \lambda _k}}\right\rangle \,dt_1 \,,
\label{probevo1}
\end{aligned}
\end{align}
where Novikov's theorem \cite{Novikov:SPJ:20:p1290:1965} has been applied. Expressions for the response functions $\delta\,\psi _i(t)/\delta\,\lambda _k(t_1)$ and $\delta\,\psi _i(t')/\delta\,\lambda _k(t_1)$ can be found by formal integration of Eq.~\eqref{sys} and iterating the formal solution. After a tedious but straightforward calculation including the computation of the response functions to 
lowest order in $(t-t_1)$ and $(t'-t_1)$ and the evaluation of several correlation integrals referring to $\chi _{kl}$ from Eq.~\eqref{noise2}, Eq.~\eqref{probevo1} can be rewritten in the limit of small correlation time $\tau$ as 
\begin{align}
\begin{aligned}
\frac{\partial}{\partial t}P_s(\boldsymbol{\psi},t;\boldsymbol{\psi '},t')=&\left\{ \mathcal{L}^0(\boldsymbol{\psi},\tau )\right. \\
																		   &\left. \quad +\exp[-(t-t')/\tau ]\,D\,\frac{\partial}{\partial \psi _i}\,\Lambda _{ik}(\boldsymbol{\psi})\,\frac{\partial}{\partial \psi '_n}\,\Lambda _{nk}(\boldsymbol{\psi '}) \right\} P_s(\boldsymbol{\psi},t;\boldsymbol{\psi '},t')\,.
\label{probevo2}
\end{aligned}
\end{align}
Thereby, transient terms and terms of the form $\propto \tau\,\exp[-(t-t')/\tau ]$ (these terms would lead to terms of 
order $\tau ^2$ in Eq.~\eqref{CFpsipsi}) have been neglected. The result is valid in the stationary case characterized by 
$t \to \infty$ and $t' \to \infty$ but finite $s = t - t'$.
In Eq.~\eqref{probevo2} $\mathcal{L}^0$ is the operator appearing in the equation for the single probability density. 
Following \cite{Hernandez-Machado:ZFPBM:52:p335:1983,DEKKER:PLA:90:p26:1982} the operator reads 
\begin{align}
\begin{aligned}
\mathcal{L}^0(\boldsymbol{\psi},\tau )= &-\frac{\partial}{\partial \psi _i}\Omega _i(\boldsymbol{\psi})+\frac{\partial}{\partial \psi _i}\Lambda _{ik}(\boldsymbol{\psi})\frac{\partial}{\partial \psi _n}\,\Biggl\{ D\,\bigl[ \Lambda _{nk}(\boldsymbol{\psi})-\tau \,M_{nk}(\boldsymbol{\psi})\bigr] \Biggr. \\ &+ \Biggl. D^2\,\tau\,\left[ K_{nkm}(\boldsymbol{\psi})\frac{\partial}{\partial \psi _l}\Lambda _{lm}(\boldsymbol{\psi})+\frac{1}{2}\Lambda _{nm}(\boldsymbol{\psi})\frac{\partial}{\partial \psi _l}K_{lkm}(\boldsymbol{\psi})\right] \Biggr\} \,,
\label{Lnull}
\end{aligned}
\end{align}
with
\begin{align}
\begin{aligned}
M_{nk} &= \Omega _r\frac{\partial \Lambda _{nk}}{\partial \psi _r}-\Lambda _{rk}\frac{\partial \Omega _n}{\partial \psi _r} \\
K_{nlk} &= \Lambda _{rk}\frac{\partial \Lambda _{nl}}{\partial \psi _r}-\frac{\partial \Lambda _{nk}}{\partial \psi _r}\Lambda _{rl}\,. 
\label{MKQ}
\end{aligned}
\end{align}
The equation of motion for the expectation value $\left\langle \psi _i\right\rangle _s$ can be evaluated from the single probability distribution in the stationary state
\be
\frac{\partial}{\partial t}P_s(\boldsymbol{\psi},t)=\mathcal{L}^0\,P_s(\boldsymbol{\psi},t)\,.
\label{singleevent}
\ee
One finds
\begin{align}
\begin{aligned}
\frac{d}{dt}\left\langle \psi _i(t)\right\rangle _s = &\left\langle \Omega _i\right\rangle _s+D\,\left\langle \frac{\partial \Lambda _{ik}}{\partial \psi _n}\bigl( \Lambda _{nk}-\tau \,M_{nk} \bigr) \right\rangle _s-D^2\,\tau \,\Biggl\{ \left\langle \frac{\partial}{\partial \psi _r} \left( \frac{\partial \Lambda _{ik}}{\partial \psi _n}K_{nkm}\right) \Lambda _{rm}\right\rangle _s \Biggr. \\
														&+\Biggl. \frac{1}{2}\,\left\langle \frac{\partial}{\partial \psi _r} \left( \frac{\partial \Lambda _{ik}}{\partial \psi _n}\Lambda _{nm}\right) K_{rkm} \right\rangle _s\Biggr\}\,. 
\label{EWpsi}
\end{aligned}
\end{align}
The knowledge of the evolution equation of the joint probability distribution $P(\boldsymbol{\psi},t;\boldsymbol{\psi '},t')$ due to 
Eqs.~\eqref{probevo2} and \eqref{Lnull} allows us to get the corresponding equation for the correlation functions.  
Following again \cite{Hernandez-Machado:ZFPBM:52:p335:1983}, it results
\begin{align}
\begin{aligned}
\frac{d}{dt}\left\langle \psi _i(t)\,\psi _j(t')\right\rangle _s = &\left\langle \Omega _i(\boldsymbol{\psi}(t))\,\psi _j(t')\right\rangle _s+D\,\left\langle \left[ \frac{\partial \Lambda _{ik}}{\partial \psi _n}\bigl( \Lambda _{nk}-\tau \,M_{nk} \bigr)\right] _t\,\psi _j(t') \right\rangle _s \\										&-D^2\,\tau \,\Biggl\{ \left\langle \left[ \frac{\partial}{\partial \psi _r} \left( \frac{\partial \Lambda _{ik}}{\partial \psi _n}K_{nkm}\right) \Lambda _{rm}\right] _t\,\psi _j(t')\right\rangle _s \Biggr. \\
														&+\Biggl. \frac{1}{2}\,\left\langle \left[ \frac{\partial}{\partial \psi _r} \left( \frac{\partial \Lambda _{ik}}{\partial \psi _n}\Lambda _{nm}\right) K_{rkm} \right] _t\,\psi _j(t')\right\rangle _s\Biggr\} \\
														&+D\,\exp\left[ -\frac{t-t'}{\tau}\right] \left\langle \Lambda _{ik}(\boldsymbol{\psi}(t))\,\Lambda _{jk}(\boldsymbol{\psi}(t'))\right\rangle _s \,,
\label{CFpsipsi}
\end{aligned}
\end{align}
where the symbol $[...]_t$ denotes the quantity $[...]$ at time $t$. As mentioned above the result is valid for $t,\,t'\to \infty$ while 
$s = t - t' > 0$ remains finite. The quantities $M_{nk}$ and $K_{klm}$ are defined in Eq.~\eqref{MKQ}. The components $\Omega _i$ and $\Lambda _{ij}$ are given in Eqs.~\eqref{omega} and \eqref{Lambda}. Performing the summation over double-indices according to Eqs.~\eqref{EWpsi} and 
\eqref{CFpsipsi} we obtain the evolution equations for the mean value and the correlation function
\be
\frac{d}{dt}\left\langle \psi _i(t)\right\rangle _s = G_{ik}\left\langle \psi _k(t)\right\rangle _s \,,
\label{EWpsi2}
\ee
and
\begin{align}
\begin{aligned}
\frac{d}{ds}\mathcal{C}_{ij}(s) = \frac{d}{ds}\left\langle \psi _i(t'+s)\,\psi _j(t')\right\rangle _s = &G_{ik}\left\langle \psi _k(t'+s)\,\psi _j(t')\right\rangle _s \\
																										&+D\,\exp\left[ -\frac{s}{\tau}\right] \left\langle \Lambda _{ik}(\boldsymbol{\psi}(t'+s))\,\Lambda _{jk}(\boldsymbol{\psi}(t'))\right\rangle _s \,.
\label{CFpsipsi2}
\end{aligned}
\end{align}
Notice, that in the steady state one gets $\mathcal{C}_{ij}(t,t')=\mathcal{C}_{ij}(s)$ with $s = t - t'$. The matrix components 
of $G_{ik}$ are given by
\be
G_{ik}=\begin{pmatrix}
						-A_1 & A_2 & 0 \\
						-A_2 & -A_1 & 0 \\
						0 & 0 & -A_3 
					\end{pmatrix}\,,
\label{MatrixG}
\ee
where 
\begin{align}
\begin{aligned}
A_1 &= -D^2\tau (6\mu ^2\alpha ^2-1)\,\xi ^4 +2\mu ^2\alpha D\tau \,\xi ^3 -D(\mu ^2\alpha ^2-2)\,\xi ^2 +\mu ^2\alpha \,\xi \\
A_2 &= \frac{1}{2}\mu \alpha D^2\tau \left( 11-3\mu ^2\alpha ^2 \right) \,\xi ^4 +\mu D \tau \left( \mu ^2\alpha ^2-1 \right) \xi ^3 +3\mu D\alpha \,\xi ^2 -\mu \,\xi \\
A_3 &= +D^2\tau \left( 3\mu ^2\alpha ^2+1\right) \xi ^4 -4\mu ^2\alpha D\tau \,\xi ^3 +2D\,\xi ^2 \,,
\label{ABH}
\end{aligned}
\end{align}
and $\xi$ is defined in Eq.~\eqref{omega}. At this point let us stress that in the case $t'=0$ the term $\propto \exp[-(t-t')/\tau]$ on the rhs. in 
Eqs.~\eqref{CFpsipsi} and \eqref{CFpsipsi2}, respectively, would vanish in the steady state, i.e. 
$$
\langle \psi _i(t'+s)\,\psi _j(t')\rangle _s \neq \langle \psi _i(s)\,\psi _j(0) \rangle _s\,.
$$
The occurrence of such a term is a strong indication for the non-stationarity of our model. An explicit calculation shows, that in general 
this inequality holds for non-stationary processes \cite{Hernandez-Machado:JoMP:25:p1066:1984}.

\section{Results}

\noindent The solution of Eq.~\eqref{EWpsi2} can be found by standard Greens function methods and Laplace transformation. As the result we find
\be
\left\langle \boldsymbol{\psi}(t)\right\rangle _s=\begin{pmatrix}
						e^{-A_1\,t}\cos( A_2\,t) & e^{-A_1\,t}\sin( A_2\,t) & 0 \\
						-e^{-A_1\,t}\sin( A_2\,t) & e^{-A_1\,t}\cos( A_2\,t) & 0 \\
						0 & 0 & e^{-A_3\,t} 
					\end{pmatrix}\cdot \left\langle \boldsymbol{\psi _0}\right\rangle _s\,,
\label{SOLpsi}
\ee
where $\left\langle \boldsymbol{\psi _0}\right\rangle _s=\left\langle \boldsymbol{\psi}(t=0)\right\rangle _s$ are the initial 
conditions. The parameters $A_1, A_3$ and $A_2$ defined in Eqs.~\eqref{ABH} play the roles of the magnon lifetime and the frequency of the spin wave at zero wave vector, respectively. As can be seen in Eq.~\eqref{ABH} all of these three parameters are affected by the correlation time $\tau$ and the strength $D$ of the random force. Moreover, the Gilbert damping parameter $\alpha$ influences the 
system as well. The solution of Eq.~\eqref{CFpsipsi2} for the correlation function in case of 
$t'=0$ is formal identical to that 
of Eq.~\eqref{SOLpsi}. The more general situation $t'\ne 0$ allows no simple analytic solution and hence the behavior of the correlation function $\mathcal{C}(s)$ is studied numerically. In order to analyze the mean values and the correlation function let us first examine 
the parameter range where physical accessible solutions exist. In the following we assume $\langle \psi _1(0)\rangle =\langle \psi _2(0)\rangle = \langle \psi _0 \rangle $ and $\langle \psi _3(0)\rangle =0$, since the solutions for $\psi _1(t)$ and $\psi _2(t)$ on the one hand and 
$\psi _3(t)$ on the other hand are decoupled in Eq.~\eqref{SOLpsi}. Therefore, spin wave solutions only exists for non-zero averages $\langle \psi _1 (t) \rangle$ and $\langle \psi _2 (t) \rangle$. The existence of such non-trivial solutions are determined in dependence on the noise parameters 
$D$ and $\tau$ and the deterministic damping parameter $\alpha$. 
Notice, that the dimensionless quantity $D=\tilde{D}/\Gamma$, i.e., $D$ is the ratio between the strength of the correlation function (Eq.~\eqref{noise1}) and the anisotropy field in the original units. The stability of spin wave solutions is guaranteed for positive parameters 
$A_1$ and $A_3$. According to Eqs.~\eqref{ABH} the phase diagrams are depicted in 
Fig.~\ref{phasespace} within the $\alpha-D$ plane for different values of the correlation time $\tau$.    
\bef
\centering
\subfigure[$\quad \tau =0$]{
				\label{psTau0}
				\includegraphics[width=0.46\linewidth ]{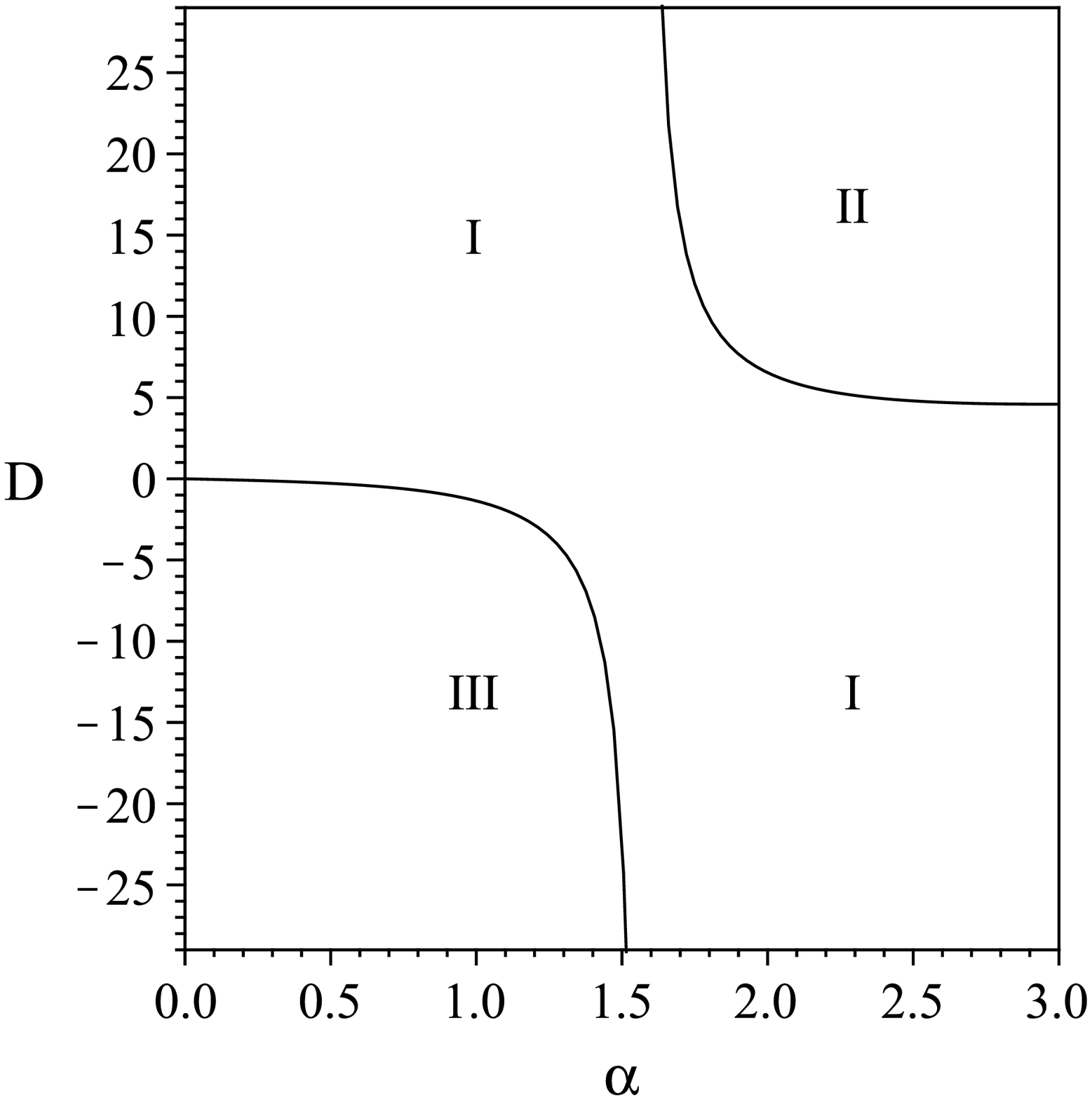}}
\hspace{0.05\linewidth}
\subfigure[$\quad \tau =0.1$]{
				\label{psTau0.1}
				\includegraphics[width=0.46\linewidth ]{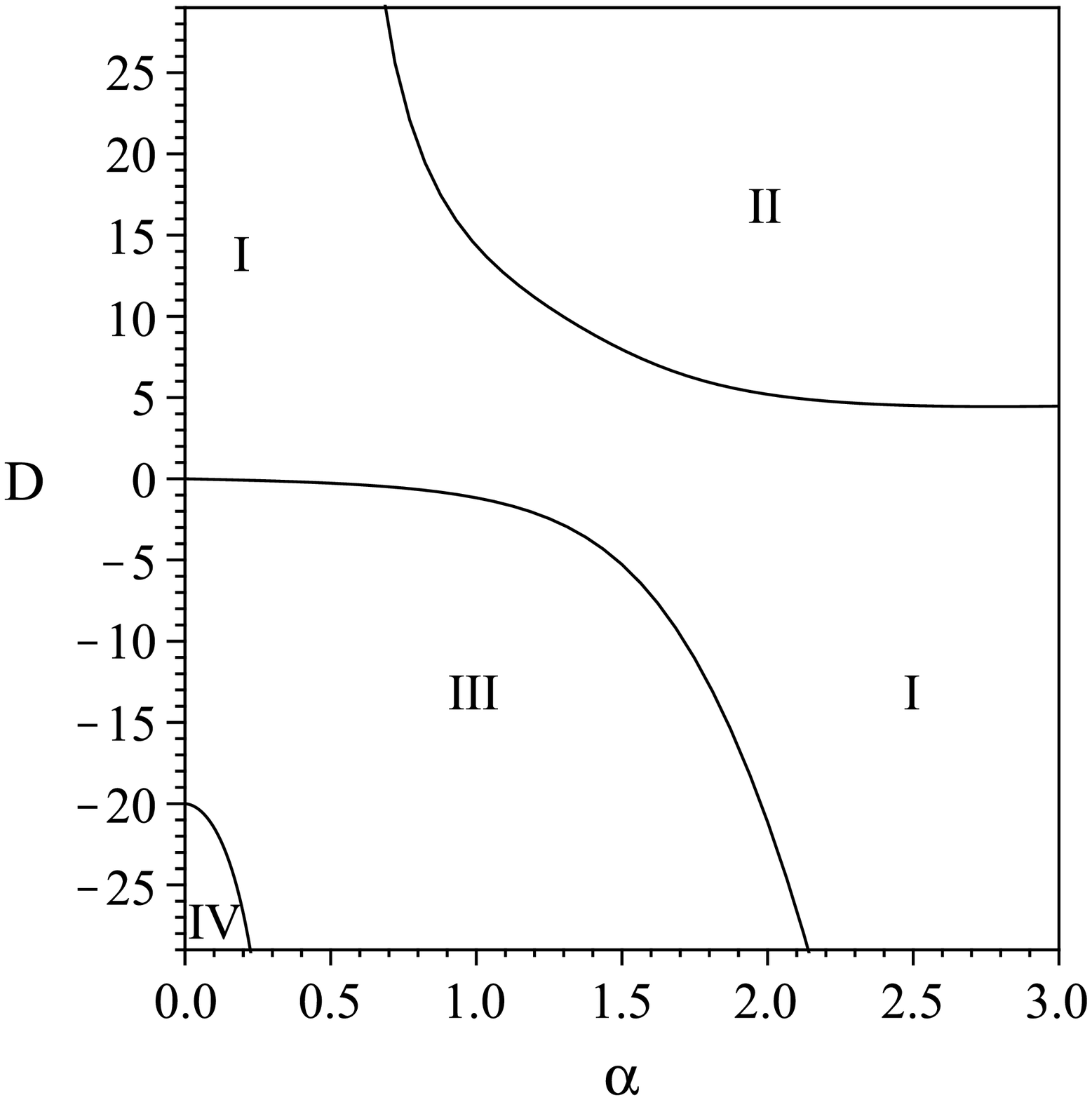}}\\[20pt]
\subfigure[$\quad \tau =1$]{
				\label{psTau1}
				\includegraphics[width=0.46\linewidth ]{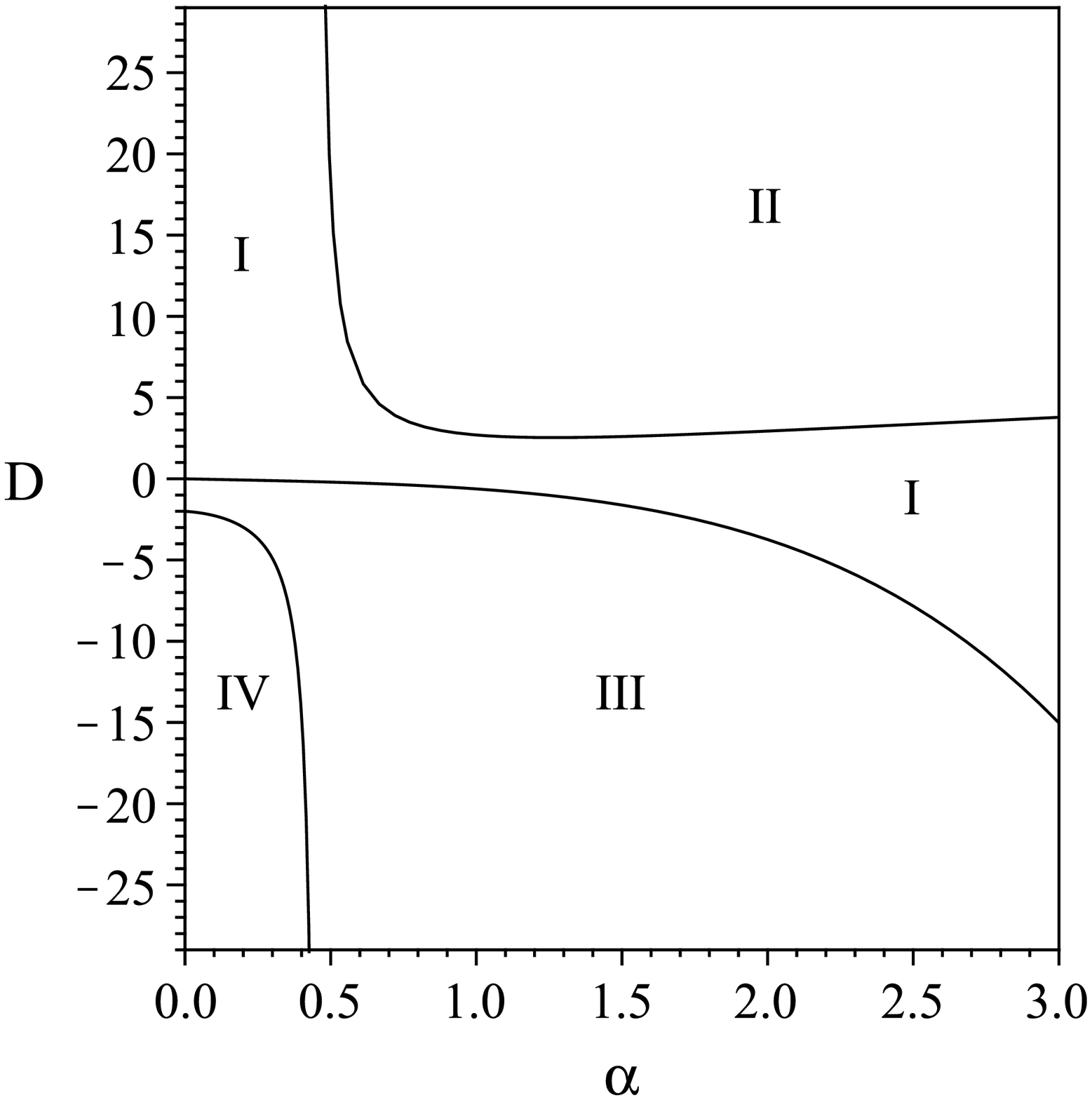}}
\hspace{0.05\linewidth}				
\subfigure[$\quad \tau =10$]{
				\label{psTau10}
				 \includegraphics[width=0.46\linewidth ]{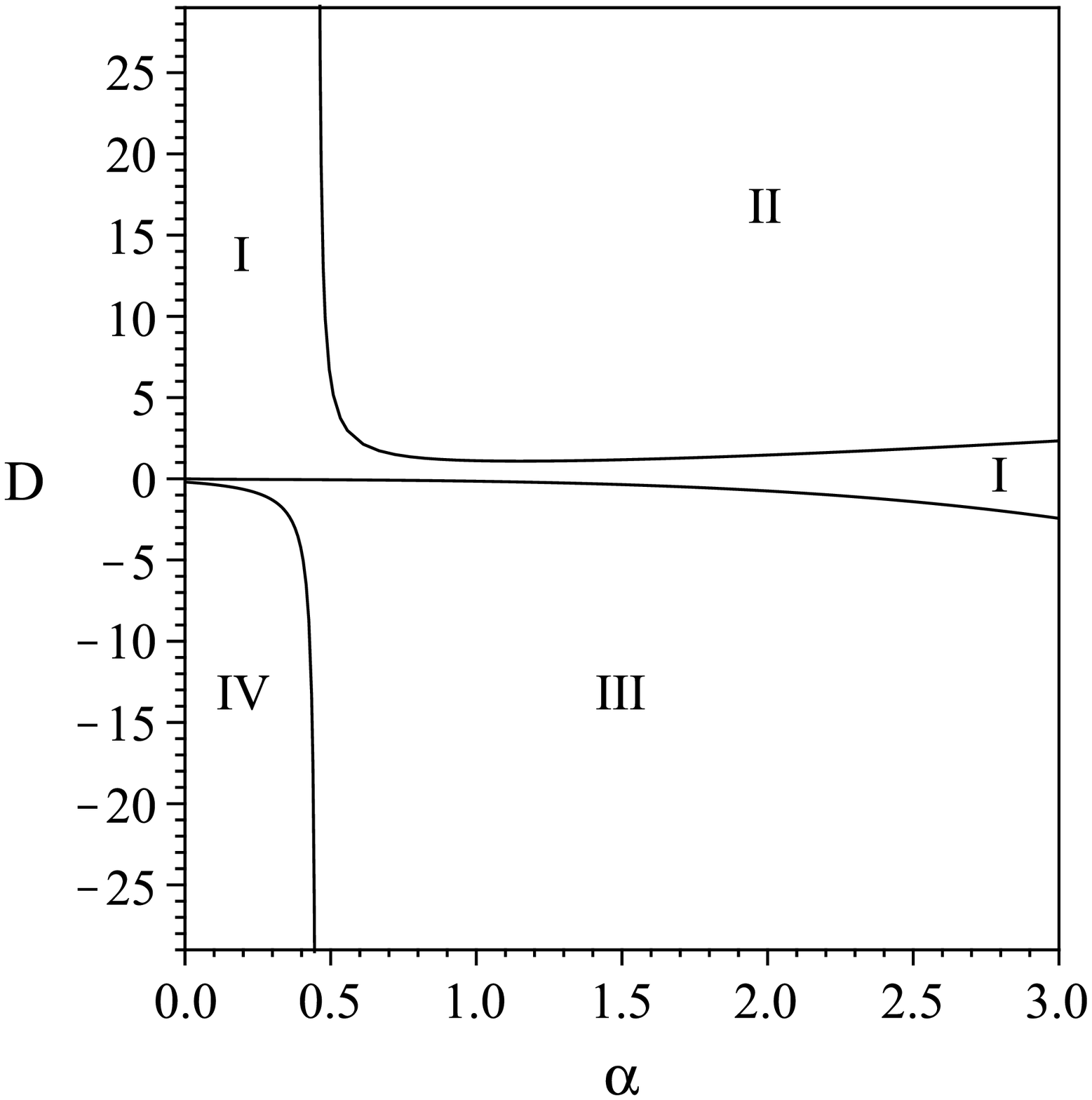}}
\caption{$\alpha-D$ plane for fixed magnetization $\mu =0.9$ and different values of $\tau$.}
\label{phasespace}
\ef
The separatrix between stable and unstable regions is determined by the condition $A_1 = 0$. 
The second condition $A_3 =0$ is irrelevant due to the imposed initial conditions. As the result 
of the stability analysis the phase space diagram is subdivided into four regions where region IV does not exist in case of $\tau =0$, see Fig.~\ref{psTau0}. For generality, we take into account both positive and negative values of $D$ indicating correlations and anti-correlations of the stochastic field. Damped spin waves are observed in the areas I and IV, whereas the sectors II and III reveal non-accessible solutions. In those regions the spin wave amplitude, 
proportional to $\exp[-A_1 t]$, tends to infinity which should not be realized, compare 
Figs.~\ref{psTau0.1}-\ref{psTau10}. Actually, a reasonable behavior is observed in regions I and IV. As visible from Fig.~\ref{phasespace} damped spin waves will always emerge for $D>0$ 
even in the limit of zero damping parameter $\alpha$ and vanishing correlation time $\tau$. This behavior is shown in Fig.~\ref{Psi1VerschAlpha}, where the evolution of $\langle \psi _1(t)\rangle$ is depicted for different values of $\alpha$. 
\bef
\includegraphics[width=0.5\linewidth ]{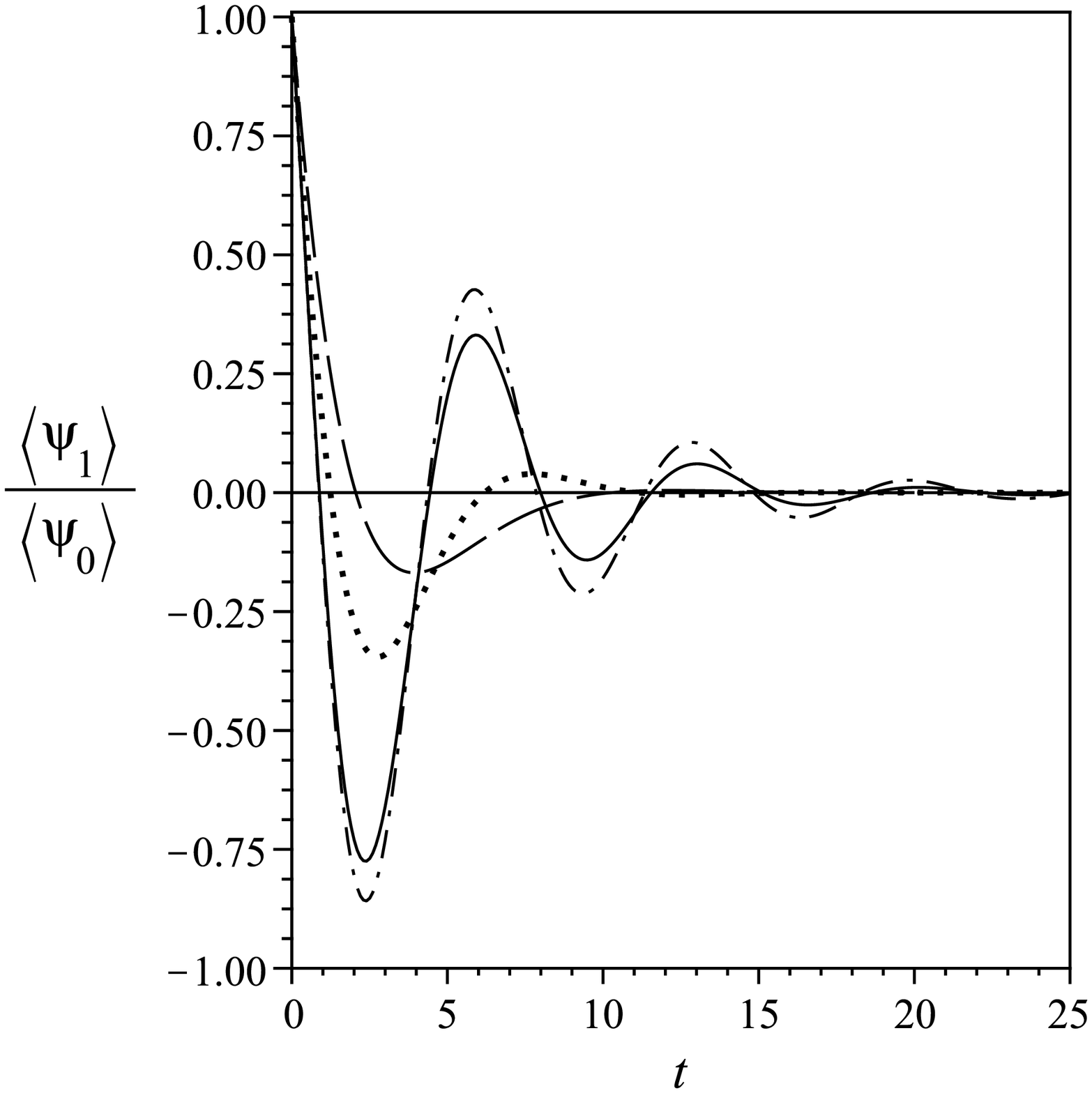}
\caption{Evolution of the mean value $\langle \psi _1(t)\rangle$, with $\mu =0.9$, $D=0.1$ and $\tau =0$. $\alpha$ varies from $0$ (dash-dotted line), $0.05$ (solid line), $0.5$ (dotted line) and $1$ (dashed line).}
\label{Psi1VerschAlpha}
\ef
As can be seen in Fig.~\ref{psTau0} the solution for $D < 0 $ is unlimited and consequently, it 
should be excluded further. Contrary to this situation, additional solutions will be developed in region IV in case of $\tau >0$ and simultaneously $\alpha = 0$, see Figs.~\ref{psTau0.1}-\ref{psTau10}. Thereby the size of area IV grows with increasing $\tau$. Likewise, the extent of region I decreases for an enhanced $\tau$. 
However, in the limit of $D=0$ and consequently for $\tau =0$, too, only damped spin waves are observed. Immediately on the separations line undamped periodic solutions will evolve, compare the sub-figures in Fig.~\ref{phasespace}. 
This remarkable effect can be traced back to the interplay between the deterministic damping and the stochastic forces. Both damping mechanism are compensated mutually which reminds of a kind of resonance phenomenon. The difference to conventional resonance behavior 
consists of the compensation of the inherent deterministic Gilbert damping and the stochastic one originated from the random field. 
This statement is emphasized by the fact that undamped periodic solutions do not develop in the absence of stochastic interferences, i.e., $D=0$. The situation might be interpreted physically as follows: the required energy that enables the system to sustain the deterministic damping mechanisms is delivered by the stochastic influences due to the interaction with the environment. To be more precise, in general, the Gilbert damping enforces the coherent alignment of the spin density along the precession axis. Contrary, the random field supports the dephasing of the 
orientation of the classical spins. Surprisingly, the model predicts the existence of a critical value $\tau =\tau _c \geq 0$ depending on $\alpha$ and $D$ which determines the onset of undamped periodic solutions. Notice, that negative values of $\tau_c$ are excluded. The critical value is
\be
\tau _c = -\frac{\left[ \mu ^2\left( \alpha ^3-D\alpha ^2+\alpha \right) +2D\right] \left( 1+\alpha ^2\right) ^2}{2D\mu ^2\left( \alpha ^3-3D\alpha ^2+\alpha \right) +D^2} \,.
\label{taucrit}
\ee
Hence, this result could imply the possibility of the cancellation of both damping processes. Examples according to the damped and the
periodic case are displayed in Fig.~\ref{picsolution}. An increasing $\tau$ favors the damping process as it is visible in 
Fig.~\ref{solpsi1difftau}. Based on estimations obtained for ferromagnetic materials \cite{Tserkovnyak:PRL:88:p117601:2002} and references
therein, the Gilbert damping parameter can range between $0.04 < \alpha < 0.22$ in thin magnetic films, whereas the bulk value for Co takes
$\alpha _b\approx 0.005$. The phase space diagram in Fig.~\ref{phasespace} offers periodic solutions only for values of $\alpha$ larger than those 
known from experiments. Therefore such periodic solutions seem to be hard to see experimentally.
\bef
\centering
\subfigure[]{
				\label{solpsi1difftau}
				\includegraphics[width=0.46\linewidth ]{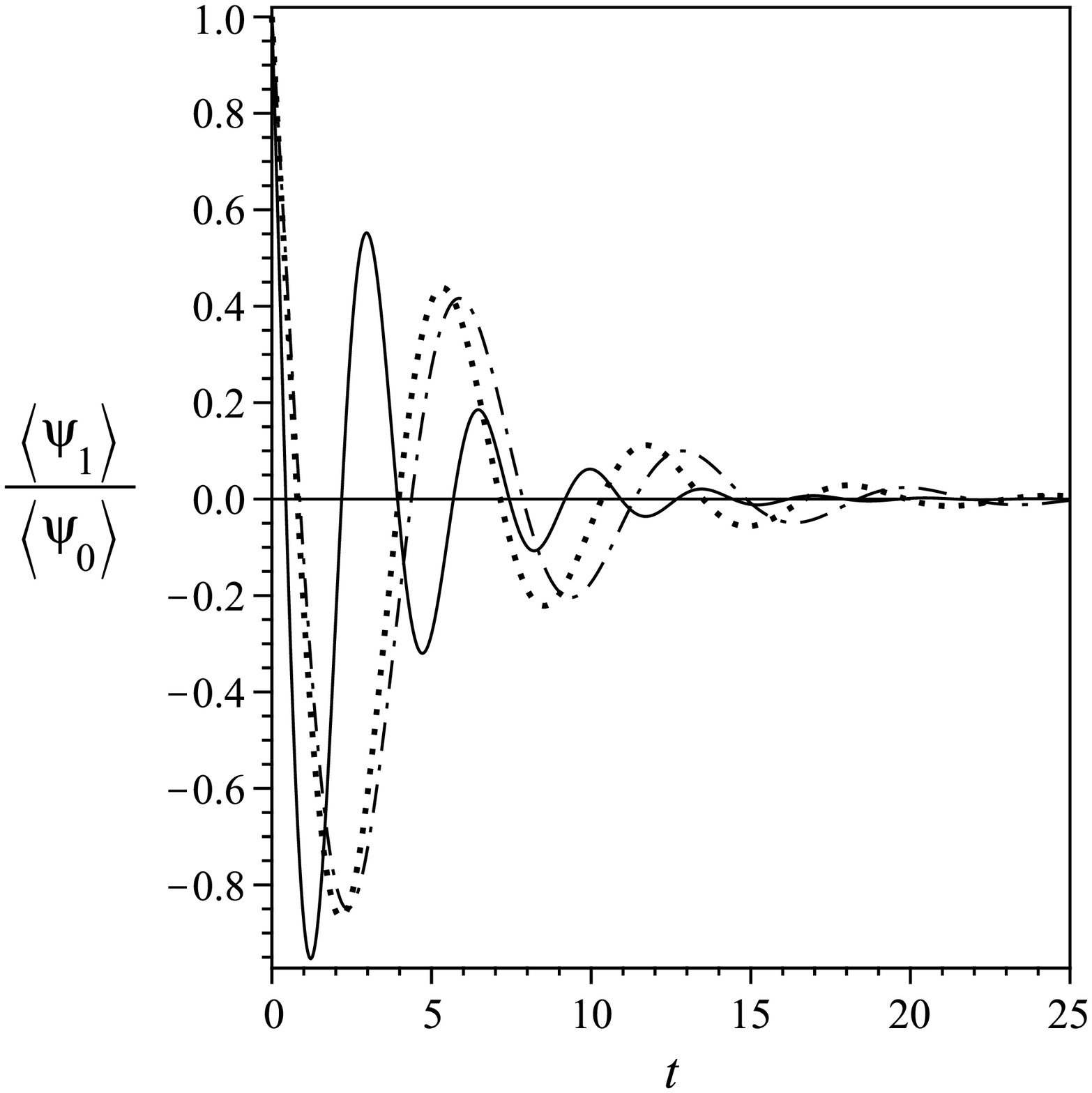}}
\hspace{0.05\linewidth}
\subfigure[]{
				\label{picsoltaucrit}
				\includegraphics[width=0.46\linewidth ]{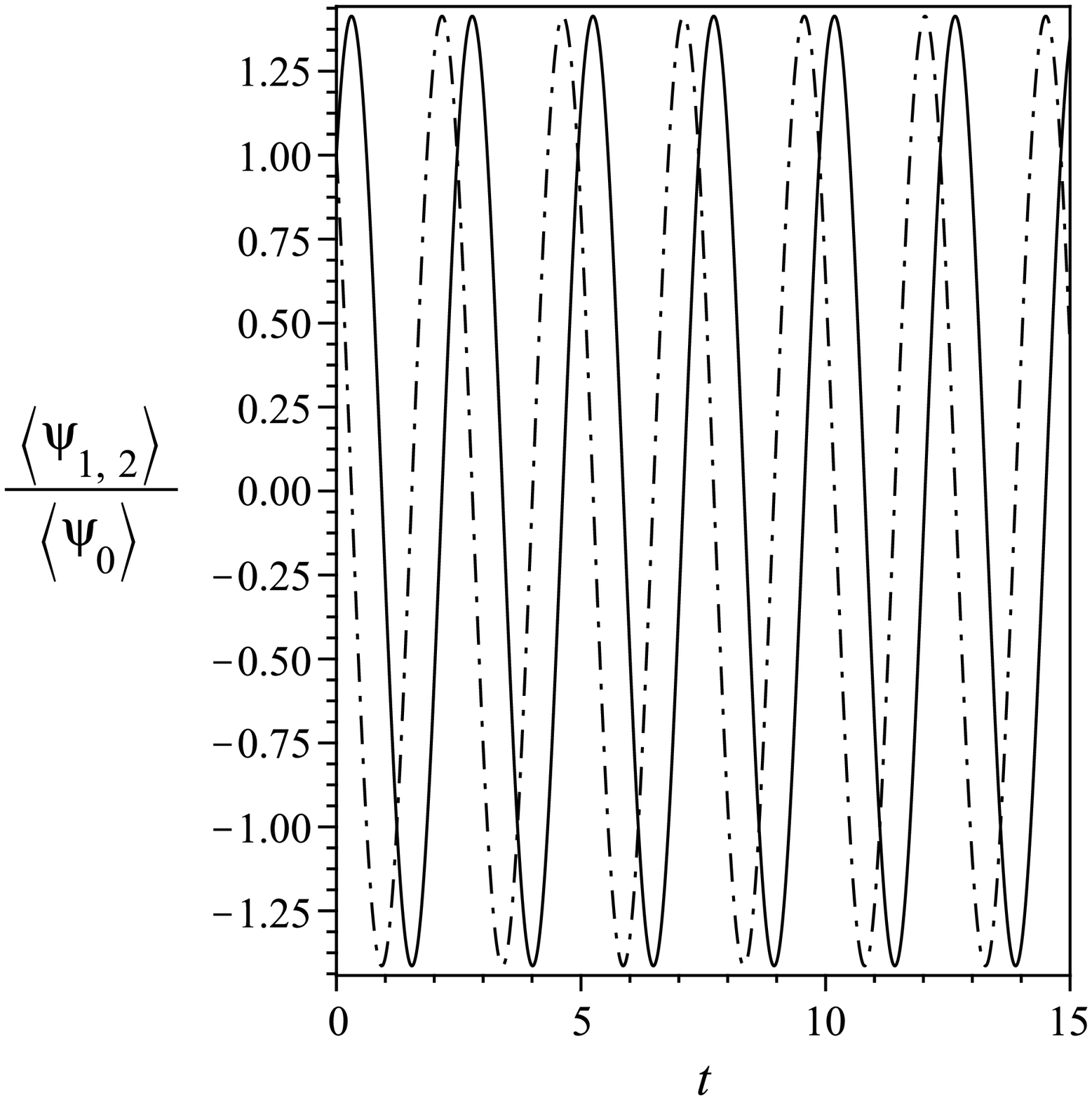}}
\caption{Evolution of the mean values $\langle \psi _{1,2}(t)\rangle$, with $\mu =0.9$. (a): $D=0.1$, $\alpha=0.005$ and $\tau$ varies from $10$ (solid line), $1$ (dotted line) and $0$ (dash-dotted line). (b): $D=2$, $\alpha=1$ and $\tau=\tau _c\approx 1.79$ (Eq.~\eqref{taucrit}). The solid line represents $\langle \psi _1\rangle$ and the dash-dotted line is $\langle \psi _2\rangle$.}
\label{picsolution}
\ef
We proceed further by analyzing the behavior of the correlation function by numerical computation of the solution of Eq.~\eqref{CFpsipsi2} with Eqs.~\eqref{MatrixG} and \eqref{ABH}. As initial values we choose \mbox{$\mathcal{C}_{ik}(t=t',t')=\mathcal{C}_{ik}(s=0)=\mathcal{C}_0$} for every combination \mbox{$i,k=\{ 1,2,3\}$}. The results are depicted in Figs.~\ref{correlfct} and \ref{CFtaucrit}. 
\bef
\centering
\subfigure[]{
				\label{C11}
				\includegraphics[width=0.46\linewidth ]{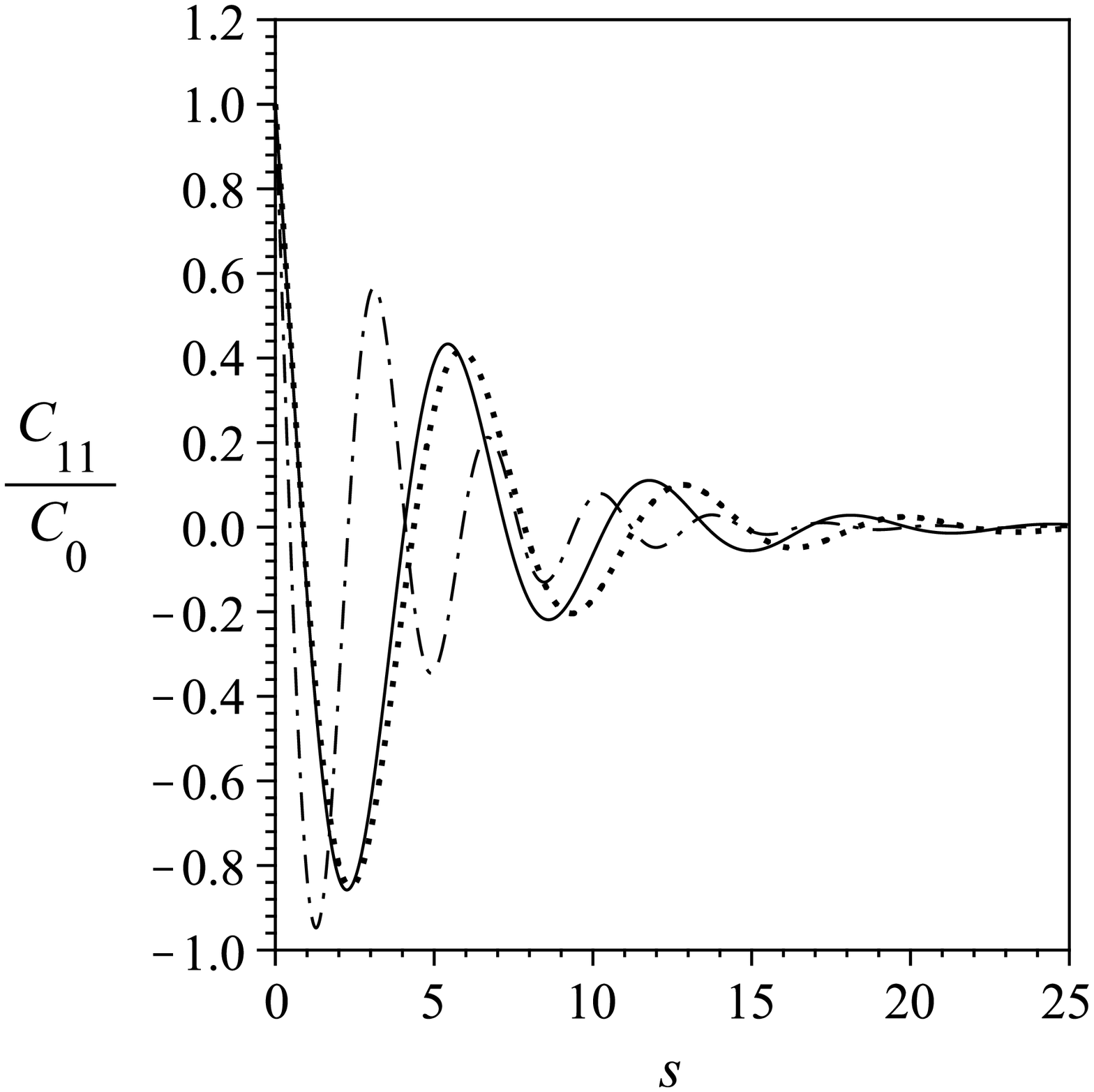}}
\hspace{0.05\linewidth}
\subfigure[]{
				\label{C12}
				\includegraphics[width=0.46\linewidth ]{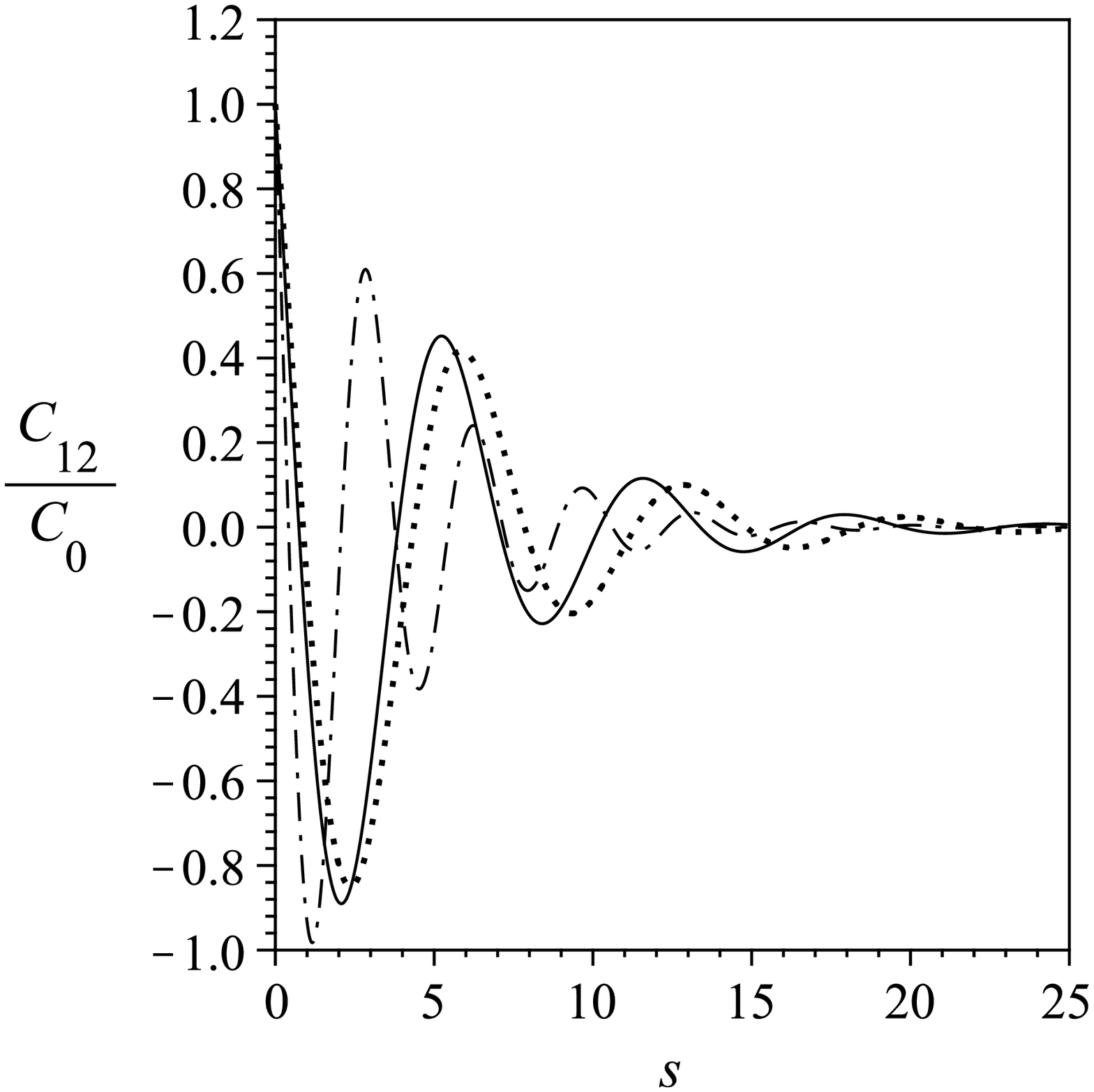}}\\[20pt]
\subfigure[]{
				\label{C13}
				\includegraphics[width=0.46\linewidth ]{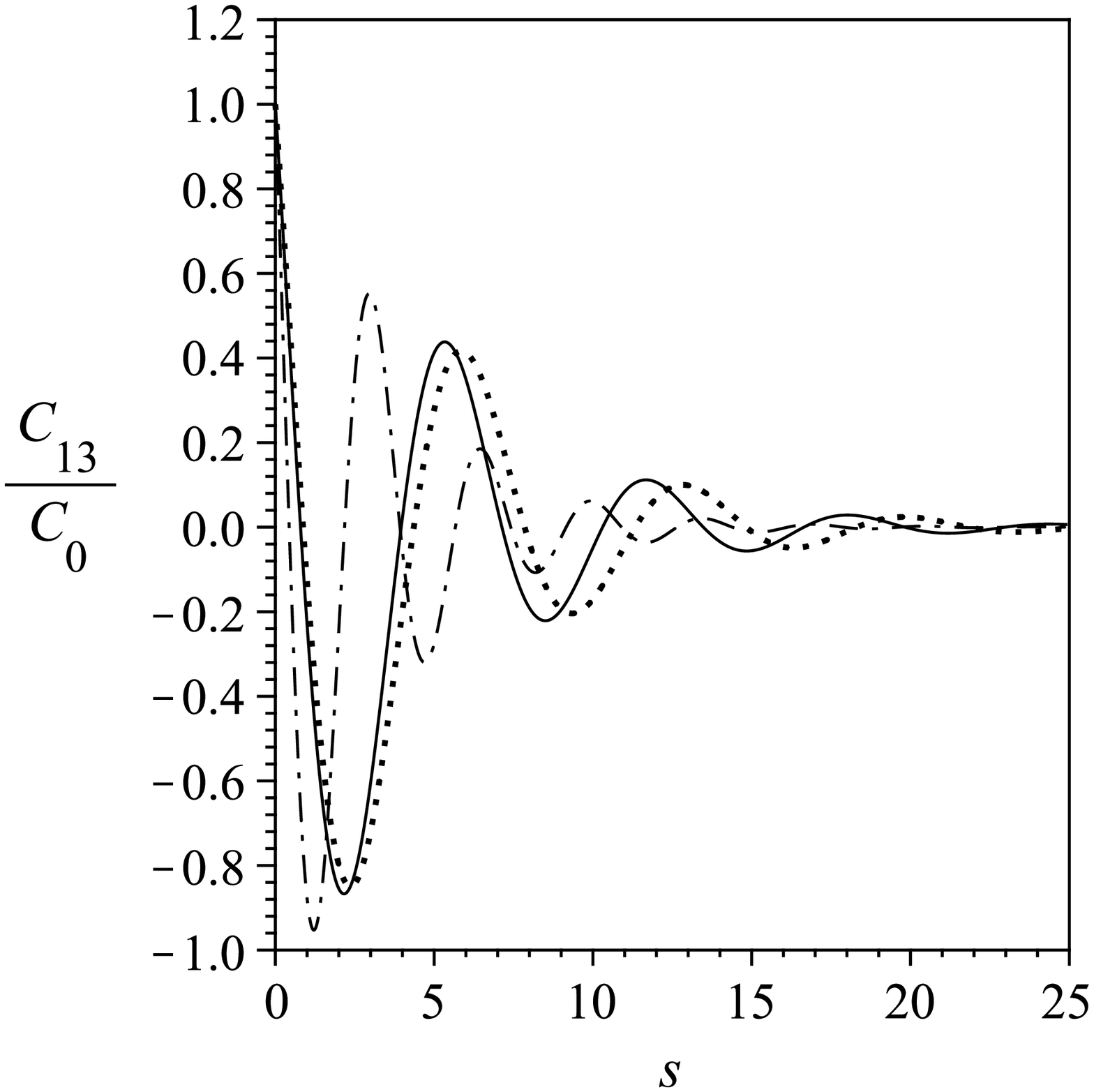}}
\hspace{0.05\linewidth}				
\subfigure[]{
				\label{C31}
				 \includegraphics[width=0.46\linewidth ]{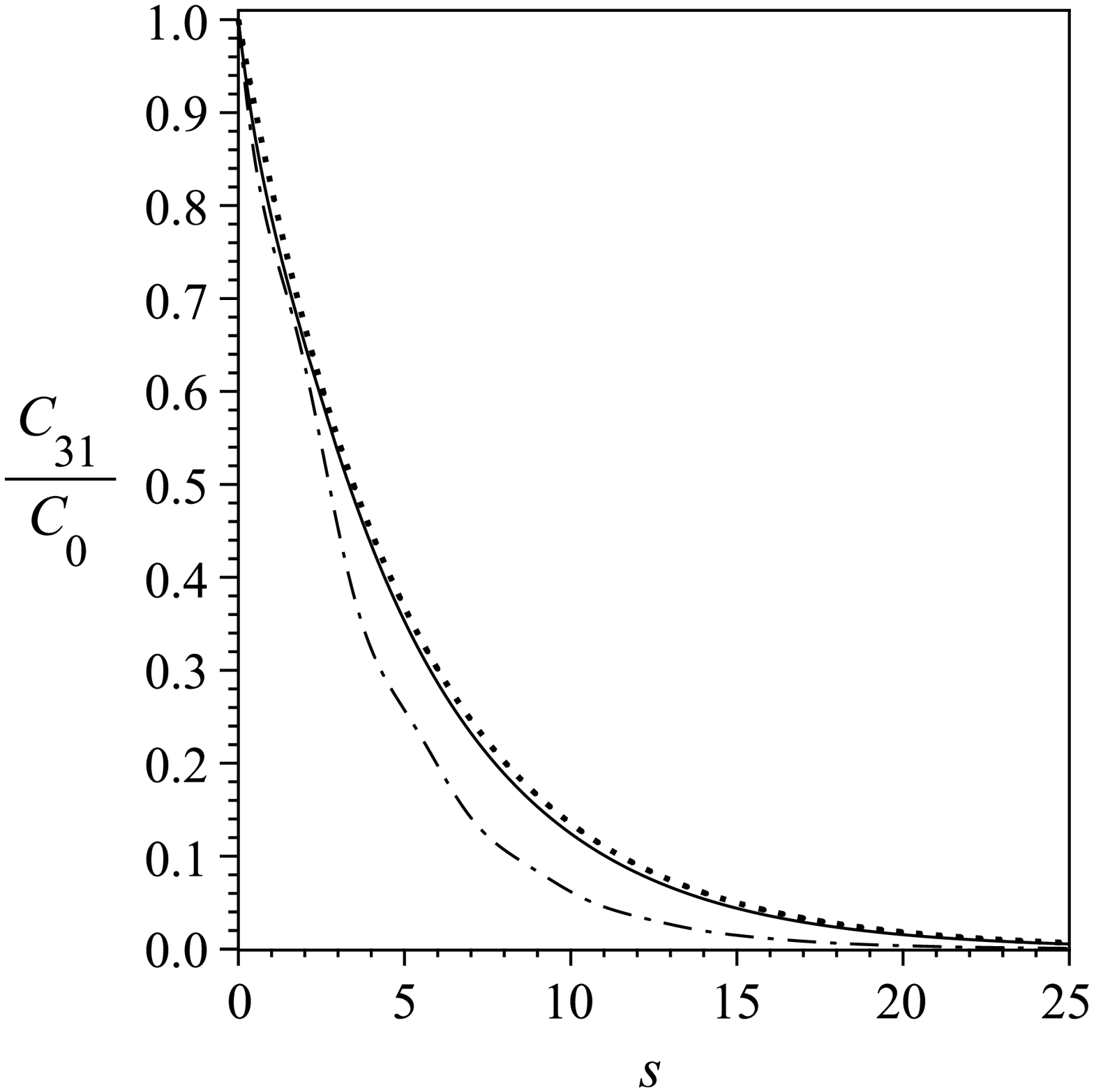}}
\caption{Correlation functions $\mathcal{C}_{ik}(s)$ for $\mu =0.9$, $D=0.1$ and $\alpha= 0.005$. $\tau$ takes $0$ (dotted line), $1$ (solid line) and $10$ (dash-dotted line).}
\label{correlfct}
\ef
Inspecting Figs.~\ref{C11}-\ref{C13} one recognizes that an enhancement of the correlation time $\tau$ leads to an increase of the oscillations within the correlation functions 
$\mathcal{C}_{1k}$, $k=\{ 1,2,3\}$. Moreover, Fig.~\ref{C31} reveals that the oscillatory 
behavior of $\mathcal{C}_{31}$ seems to be suppressed. Obviously, the decay of the correlation function is enhanced if $\tau$ growths up. The pure periodic case for $\tau =\tau _c$, 
corresponding to Fig.~\ref{picsoltaucrit}, is depicted in Fig.~\ref{CFtaucrit}. Exemplary, $\mathcal{C}_{12}$ and $\mathcal{C}_{31}$ are illustrated. The behavior of the latter is similar 
to the damped case, displayed in Fig.~\ref{C31}, unless slight oscillations occur. However, if one compares the form of $\mathcal{C}_{12}$ in Fig.~\ref{C12} and Fig.~\ref{CFtaucrit} the differences are obvious. The amplitude of the correlation function for the undamped case grows to the fourfold magnitude in comparison with $\mathcal{C}_0$, whereas the damped correlation function approaches zero. Further, a periodic behavior is shown in Fig.~\ref{CFtaucrit}, and therefore the correlation will oscillate about zero but never vanish for all \mbox{$s=t-t'>0$}. 
\bef
\includegraphics[width=0.5\linewidth ]{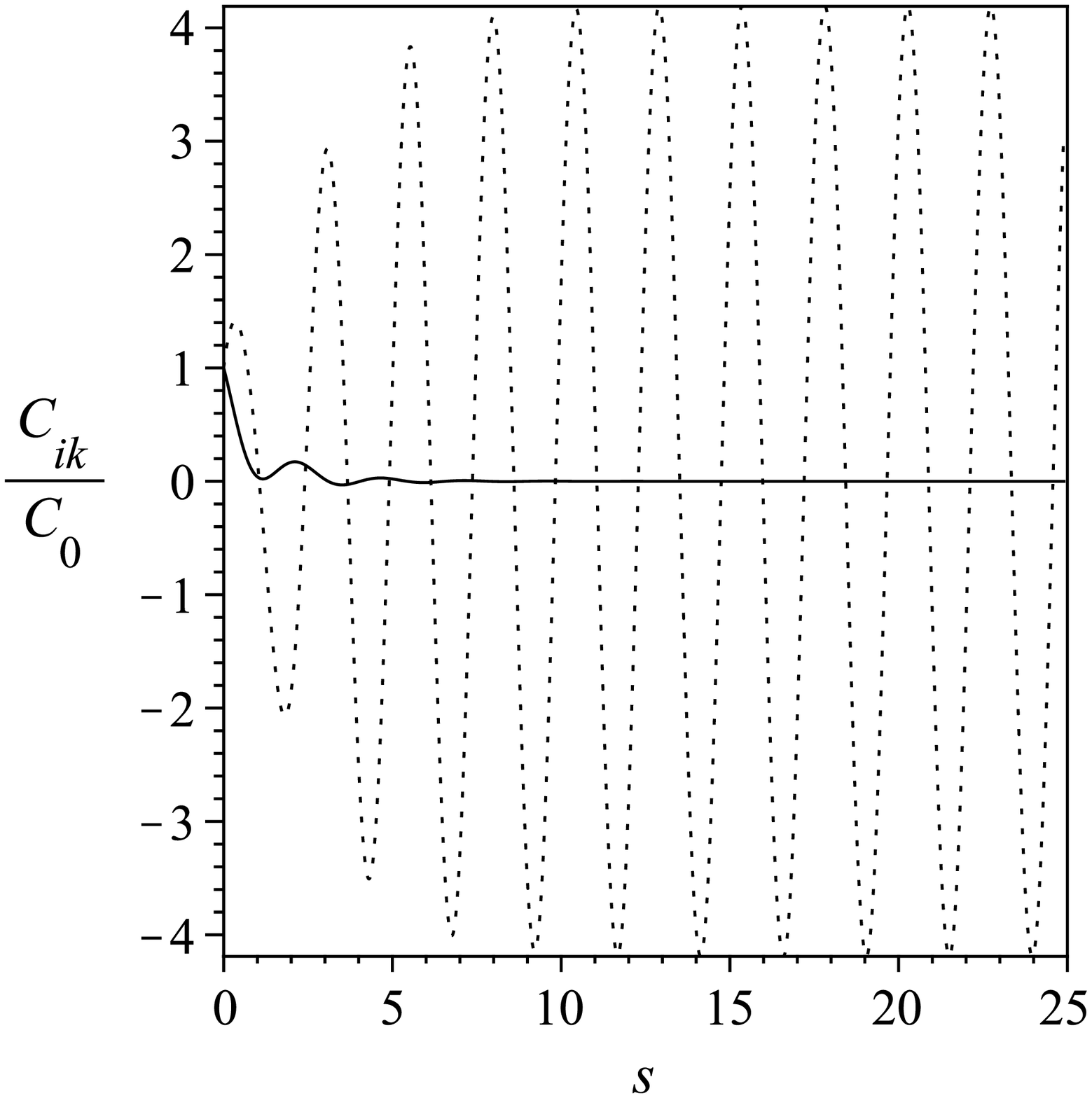}
\caption{Correlation functions $\mathcal{C}_{ik}(s)$ for $\tau =\tau _c\approx 1.79$ (Eq.~\eqref{taucrit}), $\mu =0.9$, $D=2$ and $\alpha= 1$. The dotted line represents $\mathcal{C}_{12}$ and the solid line is  $\mathcal{C}_{31}$.}
\label{CFtaucrit}
\ef

\section{Conclusions}

\noindent In this paper we have analyzed the dynamics of a classical spin model with uniaxial anisotropy. Aside from the deterministic damping due to the Landau-Lifshitz-Gilbert equation the system is subjected to an additional dissipation process by the inclusion of a stochastic field with colored noise. Both dissipation processes are able to compete leading to a more complex behavior. To study this one we derive an equation for the joint probability distribution which allows us to find the corresponding spin-spin-correlation function. This program can be fulfilled 
analytically and numerically in the spin wave approach and the small correlation time limit.  
Based on the mean value for the spin wave component and the correlation function we discuss 
the stability of the system in terms of the stochastic parameters, namely the strength of the correlated noise $D$ and the finite correlation time $\tau$, as well as the deterministic  
Gilbert damping parameter $\alpha$. The phase diagram in the $\alpha-D$ plane offers that 
the system develops stable and unstable spin wave solutions due to the interplay between 
the stochastic and the deterministic damping mechanism. So stable solutions evolve for arbitrary 
positive $D$ and moderate values of the Gilbert damping $\alpha$. Further, we find that also the finite correlation time of the stochastic field influences the evolution of the spin waves. 
In particular, the model reveals for fixed $D$ and $\alpha$ a critical value $\tau _c$ which 
characterizes the occurrence of undamped spin waves. The different situations are depicted in 
Fig.~\ref{phasespace}. Moreover, the correlation time $\tau$ affects the damped spin wave which 
can be observed in regions I and IV in the phase diagram. If the parameters $D$ and $\alpha$ 
changes within these regions, an increasing $\tau$ leads to an enhancement of the spin wave damping, cf. Fig.~\ref{solpsi1difftau}. The influence of $\tau$ on the correlation functions 
is similar as shown in Figs.~\ref{C11}-\ref{C13}. The study could be extended by the inclusion of finite wave vectors and using an approach beyond the spin wave approximation.

\begin{acknowledgments}
\noindent One of us (T.B.) is grateful to the Research Network 'Nanostructured Materials'\,, 
which is supported by the Saxony-Anhalt State, Germany.

\end{acknowledgments}

\clearpage

\newpage

\bibliography{LitNID}

\begin{thebibliography}{10}%
\makeatletter
\providecommand \@ifxundefined [1]{%
 \ifx #1\undefined \expandafter \@firstoftwo
 \else \expandafter \@secondoftwo
\fi
}%
\providecommand \@ifnum [1]{%
 \ifnum #1\expandafter \@firstoftwo
 \else \expandafter \@secondoftwo
\fi
}%
\providecommand \enquote [1]{``#1''}%
\providecommand \bibnamefont  [1]{#1}%
\providecommand \bibfnamefont [1]{#1}%
\providecommand \citenamefont [1]{#1}%
\providecommand\href[0]{\@sanitize\@href}%
\providecommand\@href[1]{\endgroup\@@startlink{#1}\endgroup\@@href}%
\providecommand\@@href[1]{#1\@@endlink}%
\providecommand \@sanitize [0]{\begingroup\catcode`\&12\catcode`\#12\relax}%
\@ifxundefined \pdfoutput {\@firstoftwo}{%
 \@ifnum{\z@=\pdfoutput}{\@firstoftwo}{\@secondoftwo}%
}{%
 \providecommand\@@startlink[1]{\leavevmode\special{html:<a href="#1">}}%
 \providecommand\@@endlink[0]{\special{html:</a>}}%
}{%
 \providecommand\@@startlink[1]{%
  \leavevmode
  \pdfstartlink
   attr{/Border[0 0 1 ]/H/I/C[0 1 1]}%
   user{/Subtype/Link/A<</Type/Action/S/URI/URI(#1)>>}%
  \relax
 }%
 \providecommand\@@endlink[0]{\pdfendlink}%
}%
\providecommand \url  [0]{\begingroup\@sanitize \@url }%
\providecommand \@url [1]{\endgroup\@href {#1}{\urlprefix}}%
\providecommand \urlprefix [0]{URL }%
\providecommand \Eprint[0]{\href }%
\@ifxundefined \urlstyle {%
  \providecommand \doi [1]{doi:\discretionary{}{}{}#1}%
}{%
  \providecommand \doi [0]{doi:\discretionary{}{}{}\begingroup
  \urlstyle{rm}\Url }%
}%
\providecommand \doibase [0]{http://dx.doi.org/}%
\providecommand \Doi[1]{\href{\doibase#1}}%
\providecommand \bibAnnote [3]{%
  \BibitemShut{#1}%
  \begin{quotation}\noindent
    \textsc{Key:}\ #2\\\textsc{Annotation:}\ #3%
  \end{quotation}%
}%
\providecommand \bibAnnoteFile [2]{%
  \IfFileExists{#2}{\bibAnnote {#1} {#2} {\input{#2}}}{}%
}%
\providecommand \typeout [0]{\immediate \write \m@ne }%
\providecommand \selectlanguage [0]{\@gobble}%
\providecommand \bibinfo [0]{\@secondoftwo}%
\providecommand \bibfield [0]{\@secondoftwo}%
\providecommand \translation [1]{[#1]}%
\providecommand \BibitemOpen[0]{}%
\providecommand \bibitemStop [0]{}%
\providecommand \bibitemNoStop [0]{.\EOS\space}%
\providecommand \EOS [0]{\spacefactor3000\relax}%
\providecommand \BibitemShut [1]{\csname bibitem#1\endcsname}%
\bibitem{Landau:ElecContMed:Book:1989}%
  \BibitemOpen
  \bibfield{author}{%
  \bibinfo {author} {\bibfnamefont{L.~D.}\ \bibnamefont{Landau}}, \bibinfo
  {author} {\bibfnamefont{E.}~\bibnamefont{Lifshitz}},\ and\ \bibinfo {author}
  {\bibfnamefont{L.}~\bibnamefont{Pitaevskii}},\ }%
  \emph{\bibinfo {title} {Electrodynamics of continuous media}}\ (\bibinfo
  {publisher} {Pergamon Press},\ \bibinfo {address} {Oxford},\ \bibinfo {year}
  {1989})%
  \bibAnnoteFile{NoStop}{Landau:ElecContMed:Book:1989}%
\bibitem{Landau:ZdS:8:p153:1935}%
  \BibitemOpen
  \bibfield{author}{%
  \bibinfo {author} {\bibfnamefont{L.}~\bibnamefont{Landau}}\ and\ \bibinfo
  {author} {\bibfnamefont{E.}~\bibnamefont{Lifshitz}},\ }%
  \bibfield{journal}{%
  \bibinfo {journal} {Zeitschr. d. Sowj.}\ }%
  \textbf{\bibinfo {volume} {8}},\ \bibinfo {pages} {153} (\bibinfo {year}
  {1935})%
  \bibAnnoteFile{NoStop}{Landau:ZdS:8:p153:1935}%
\bibitem{Tserkovnyak:RoMP:77:p1375:2005}%
  \BibitemOpen
  \bibfield{author}{%
  \bibinfo {author} {\bibfnamefont{Y.}~\bibnamefont{Tserkovnyak}}, \bibinfo
  {author} {\bibfnamefont{A.}~\bibnamefont{Brataas}}, \bibinfo {author}
  {\bibfnamefont{G.~E.~W.}\ \bibnamefont{Bauer}},\ and\ \bibinfo {author}
  {\bibfnamefont{B.~I.}\ \bibnamefont{Halperin}},\ }%
  \bibfield{journal}{%
  \bibinfo {journal} {Rev. Mod. Phys.}\ }%
  \textbf{\bibinfo {volume} {77}},\ \bibinfo {pages} {1375} (\bibinfo {year}
  {2005})%
  \bibAnnoteFile{NoStop}{Tserkovnyak:RoMP:77:p1375:2005}%
\bibitem{Sukhov:JoPM:20:p125226:2008}%
  \BibitemOpen
  \bibfield{author}{%
  \bibinfo {author} {\bibfnamefont{A.}~\bibnamefont{Sukhov}}\ and\ \bibinfo
  {author} {\bibfnamefont{J.}~\bibnamefont{Berakdar}},\ }%
  \bibfield{journal}{%
  \bibinfo {journal} {J. Phys. - Cond. Mat.}\ }%
  \textbf{\bibinfo {volume} {20}},\ \bibinfo {pages} {125226} (\bibinfo {year}
  {2008})%
  \bibAnnoteFile{NoStop}{Sukhov:JoPM:20:p125226:2008}%
\bibitem{Slonczewski:JoMaMM:159:p1:1996}%
  \BibitemOpen
  \bibfield{author}{%
  \bibinfo {author} {\bibfnamefont{J.~C.}\ \bibnamefont{Slonczewski}},\ }%
  \bibfield{journal}{%
  \bibinfo {journal} {J. Magn. and Mag. Mat.}\ }%
  \textbf{\bibinfo {volume} {159}},\ \bibinfo {pages} {L1} (\bibinfo {year}
  {1996})%
  \bibAnnoteFile{NoStop}{Slonczewski:JoMaMM:159:p1:1996}%
\bibitem{Berger:PRB:54:p9353:1996}%
  \BibitemOpen
  \bibfield{author}{%
  \bibinfo {author} {\bibfnamefont{L.}~\bibnamefont{Berger}},\ }%
  \bibfield{journal}{%
  \bibinfo {journal} {Phys. Rev. B}\ }%
  \textbf{\bibinfo {volume} {54}},\ \bibinfo {pages} {9353} (\bibinfo {year}
  {1996})%
  \bibAnnoteFile{NoStop}{Berger:PRB:54:p9353:1996}%
\bibitem{Urazhdin:PRB:78:p60405:2008}%
  \BibitemOpen
  \bibfield{author}{%
  \bibinfo {author} {\bibfnamefont{S.}~\bibnamefont{Urazhdin}},\ }%
  \bibfield{journal}{%
  \bibinfo {journal} {Phys. Rev. B}\ }%
  \textbf{\bibinfo {volume} {78}},\ \bibinfo {pages} {060405} (\bibinfo {year}
  {2008})%
  \bibAnnoteFile{NoStop}{Urazhdin:PRB:78:p60405:2008}%
\bibitem{Kruger:PRB:75:p54421:2007}%
  \BibitemOpen
  \bibfield{author}{%
  \bibinfo {author} {\bibfnamefont{B.}~\bibnamefont{Krüger}}, \bibinfo {author}
  {\bibfnamefont{D.}~\bibnamefont{Pfannkuche}}, \bibinfo {author}
  {\bibfnamefont{M.}~\bibnamefont{Bolte}}, \bibinfo {author}
  {\bibfnamefont{G.}~\bibnamefont{Meier}},\ and\ \bibinfo {author}
  {\bibfnamefont{U.}~\bibnamefont{Merkt}},\ }%
  \bibfield{journal}{%
  \bibinfo {journal} {Phys. Rev. B}\ }%
  \textbf{\bibinfo {volume} {75}},\ \bibinfo {pages} {054421} (\bibinfo {year}
  {2007})%
  \bibAnnoteFile{NoStop}{Kruger:PRB:75:p54421:2007}%
\bibitem{Usadel:PRB:73:p212405:2006}%
  \BibitemOpen
  \bibfield{author}{%
  \bibinfo {author} {\bibfnamefont{K.~D.}\ \bibnamefont{Usadel}},\ }%
  \bibfield{journal}{%
  \bibinfo {journal} {Phys. Rev. B}\ }%
  \textbf{\bibinfo {volume} {73}},\ \bibinfo {pages} {212405} (\bibinfo {year}
  {2006})%
  \bibAnnoteFile{NoStop}{Usadel:PRB:73:p212405:2006}%
\bibitem{Gilbert:ITOM:40:p3443:2004}%
  \BibitemOpen
  \bibfield{author}{%
  \bibinfo {author} {\bibfnamefont{T.~L.}\ \bibnamefont{Gilbert}},\ }%
  \bibfield{journal}{%
  \bibinfo {journal} {IEEE Trans. Magn.}\ }%
  \textbf{\bibinfo {volume} {40}},\ \bibinfo {pages} {3443} (\bibinfo {year}
  {2004})%
  \bibAnnoteFile{NoStop}{Gilbert:ITOM:40:p3443:2004}%
\bibitem{Hickey:PRL:102:p137601:2009}%
  \BibitemOpen
  \bibfield{author}{%
  \bibinfo {author} {\bibfnamefont{M.~C.}\ \bibnamefont{Hickey}}\ and\ \bibinfo
  {author} {\bibfnamefont{J.~S.}\ \bibnamefont{Moodera}},\ }%
  \bibfield{journal}{%
  \bibinfo {journal} {Phys. Rev. Lett.}\ }%
  \textbf{\bibinfo {volume} {102}},\ \bibinfo {pages} {137601} (\bibinfo {year}
  {2009})%
  \bibAnnoteFile{NoStop}{Hickey:PRL:102:p137601:2009}%
\bibitem{Zhang:PRL:102:p86601:2009}%
  \BibitemOpen
  \bibfield{author}{%
  \bibinfo {author} {\bibfnamefont{S.~F.}\ \bibnamefont{Zhang}}\ and\ \bibinfo
  {author} {\bibfnamefont{S.~S.~L.}\ \bibnamefont{Zhang}},\ }%
  \bibfield{journal}{%
  \bibinfo {journal} {Phys. Rev. Lett.}\ }%
  \textbf{\bibinfo {volume} {102}},\ \bibinfo {pages} {086601} (\bibinfo {year}
  {2009})%
  \bibAnnoteFile{NoStop}{Zhang:PRL:102:p86601:2009}%
\bibitem{Trimper:PRB:76:p94108:2007}%
  \BibitemOpen
  \bibfield{author}{%
  \bibinfo {author} {\bibfnamefont{S.}~\bibnamefont{Trimper}}, \bibinfo
  {author} {\bibfnamefont{T.}~\bibnamefont{Michael}},\ and\ \bibinfo {author}
  {\bibfnamefont{J.~M.}\ \bibnamefont{Wesselinowa}},\ }%
  \bibfield{journal}{%
  \bibinfo {journal} {Phys. Rev. B}\ }%
  \textbf{\bibinfo {volume} {76}},\ \bibinfo {pages} {094108} (\bibinfo {year}
  {2007})%
  \bibAnnoteFile{NoStop}{Trimper:PRB:76:p94108:2007}%
\bibitem{Foros:PRB:79:p214407:2009}%
  \BibitemOpen
  \bibfield{author}{%
  \bibinfo {author} {\bibfnamefont{J.}~\bibnamefont{Foros}}, \bibinfo {author}
  {\bibfnamefont{A.}~\bibnamefont{Brataas}}, \bibinfo {author}
  {\bibfnamefont{G.~E.~W.}\ \bibnamefont{Bauer}},\ and\ \bibinfo {author}
  {\bibfnamefont{Y.}~\bibnamefont{Tserkovnyak}},\ }%
  \bibfield{journal}{%
  \bibinfo {journal} {Phys. Rev. B}\ }%
  \textbf{\bibinfo {volume} {79}},\ \bibinfo {pages} {214407} (\bibinfo {year}
  {2009})%
  \bibAnnoteFile{NoStop}{Foros:PRB:79:p214407:2009}%
\bibitem{Chudnovskiy:PRL:101:p66601:2008}%
  \BibitemOpen
  \bibfield{author}{%
  \bibinfo {author} {\bibfnamefont{A.~L.}\ \bibnamefont{Chudnovskiy}}, \bibinfo
  {author} {\bibfnamefont{J.}~\bibnamefont{Swiebodzinski}},\ and\ \bibinfo
  {author} {\bibfnamefont{A.}~\bibnamefont{Kamenev}},\ }%
  \bibfield{journal}{%
  \bibinfo {journal} {Phys. Rev. Lett.}\ }%
  \textbf{\bibinfo {volume} {101}},\ \bibinfo {pages} {066601} (\bibinfo {year}
  {2008})%
  \bibAnnoteFile{NoStop}{Chudnovskiy:PRL:101:p66601:2008}%
\bibitem{Basko:PRB:79:p64418:2009}%
  \BibitemOpen
  \bibfield{author}{%
  \bibinfo {author} {\bibfnamefont{D.~M.}\ \bibnamefont{Basko}}\ and\ \bibinfo
  {author} {\bibfnamefont{M.~G.}\ \bibnamefont{Vavilov}},\ }%
  \bibfield{journal}{%
  \bibinfo {journal} {Phys. Rev. B}\ }%
  \textbf{\bibinfo {volume} {79}},\ \bibinfo {pages} {064418} (\bibinfo {year}
  {2009})%
  \bibAnnoteFile{NoStop}{Basko:PRB:79:p64418:2009}%
\bibitem{Denisov:PRB:75:p184432:2007}%
  \BibitemOpen
  \bibfield{author}{%
  \bibinfo {author} {\bibfnamefont{S.~I.}\ \bibnamefont{Denisov}}, \bibinfo
  {author} {\bibfnamefont{K.}~\bibnamefont{Sakmann}}, \bibinfo {author}
  {\bibfnamefont{P.}~\bibnamefont{Talkner}},\ and\ \bibinfo {author}
  {\bibfnamefont{P.}~\bibnamefont{Hänggi}},\ }%
  \bibfield{journal}{%
  \bibinfo {journal} {Phys. Rev. B}\ }%
  \textbf{\bibinfo {volume} {75}},\ \bibinfo {pages} {184432} (\bibinfo {year}
  {2007})%
  \bibAnnoteFile{NoStop}{Denisov:PRB:75:p184432:2007}%
\bibitem{Daniel:PAMaIA:120:p125:1983}%
  \BibitemOpen
  \bibfield{author}{%
  \bibinfo {author} {\bibfnamefont{M.}~\bibnamefont{Daniel}}\ and\ \bibinfo
  {author} {\bibfnamefont{M.}~\bibnamefont{Lakshmanan}},\ }%
  \bibfield{journal}{%
  \bibinfo {journal} {Physica A}\ }%
  \textbf{\bibinfo {volume} {120}},\ \bibinfo {pages} {125} (\bibinfo {year}
  {1983})%
  \bibAnnoteFile{NoStop}{Daniel:PAMaIA:120:p125:1983}%
\bibitem{Lakshmanan:PRL:53:p2497:1984}%
  \BibitemOpen
  \bibfield{author}{%
  \bibinfo {author} {\bibfnamefont{M.}~\bibnamefont{Lakshmanan}}\ and\ \bibinfo
  {author} {\bibfnamefont{K.}~\bibnamefont{Nakamura}},\ }%
  \bibfield{journal}{%
  \bibinfo {journal} {Phys. Rev. Lett.}\ }%
  \textbf{\bibinfo {volume} {53}},\ \bibinfo {pages} {2497} (\bibinfo {year}
  {1984})%
  \bibAnnoteFile{NoStop}{Lakshmanan:PRL:53:p2497:1984}%
\bibitem{Bar'Yakhtar:DynTopMagSol:Book:1994}%
  \BibitemOpen
  \bibfield{author}{%
  \bibinfo {author} {\bibfnamefont{V.~G.}\ \bibnamefont{Bar'Yakhtar}}, \bibinfo
  {author} {\bibfnamefont{M.~V.}\ \bibnamefont{Chetkin}}, \bibinfo {author}
  {\bibfnamefont{B.~A.}\ \bibnamefont{Ivanov}},\ and\ \bibinfo {author}
  {\bibfnamefont{S.~N.}\ \bibnamefont{Gadetskii}},\ }%
  \emph{\bibinfo {title} {Dynamics of Topological Magnetic Solitons: Experiment
  and Theory (Springer Tracts in Modern Physics)}}\ (\bibinfo {publisher}
  {Springer},\ \bibinfo {year} {1994})%
  \bibAnnoteFile{NoStop}{Bar'Yakhtar:DynTopMagSol:Book:1994}%
\bibitem{Lakshmanan:PA:84:p577:1976}%
  \BibitemOpen
  \bibfield{author}{%
  \bibinfo {author} {\bibfnamefont{M.}~\bibnamefont{Lakshmanan}}, \bibinfo
  {author} {\bibfnamefont{T.~W.}\ \bibnamefont{Ruijgrok}},\ and\ \bibinfo
  {author} {\bibfnamefont{C.~J.}\ \bibnamefont{Thompson}},\ }%
  \bibfield{journal}{%
  \bibinfo {journal} {Physica A}\ }%
  \textbf{\bibinfo {volume} {84}},\ \bibinfo {pages} {577} (\bibinfo {year}
  {1976})%
  \bibAnnoteFile{NoStop}{Lakshmanan:PA:84:p577:1976}%
\bibitem{Kosevich:PRSoPL:194:p117:1990}%
  \BibitemOpen
  \bibfield{author}{%
  \bibinfo {author} {\bibfnamefont{A.~M.}\ \bibnamefont{Kosevich}}, \bibinfo
  {author} {\bibfnamefont{B.~A.}\ \bibnamefont{Ivanov}},\ and\ \bibinfo
  {author} {\bibfnamefont{A.~S.}\ \bibnamefont{Kovalev}},\ }%
  \bibfield{journal}{%
  \bibinfo {journal} {Phys. Rep.}\ }%
  \textbf{\bibinfo {volume} {194}},\ \bibinfo {pages} {117} (\bibinfo {year}
  {1990})%
  \bibAnnoteFile{NoStop}{Kosevich:PRSoPL:194:p117:1990}%
\bibitem{Hernandez-Machado:JoMP:25:p1066:1984}%
  \BibitemOpen
  \bibfield{author}{%
  \bibinfo {author} {\bibfnamefont{A.}~\bibnamefont{Hernandez-Machado}}\ and\
  \bibinfo {author} {\bibfnamefont{M.}~\bibnamefont{San~Miguel}},\ }%
  \bibfield{journal}{%
  \bibinfo {journal} {J. Math. Phys.}\ }%
  \textbf{\bibinfo {volume} {25}},\ \bibinfo {pages} {1066} (\bibinfo {year}
  {1984})%
  \bibAnnoteFile{NoStop}{Hernandez-Machado:JoMP:25:p1066:1984}%
\bibitem{Hernandez-Machado:ZFPBM:52:p335:1983}%
  \BibitemOpen
  \bibfield{author}{%
  \bibinfo {author} {\bibfnamefont{A.}~\bibnamefont{Hernandez-Machado}},
  \bibinfo {author} {\bibfnamefont{J.~M.}\ \bibnamefont{Sancho}}, \bibinfo
  {author} {\bibfnamefont{M.}~\bibnamefont{San~Miguel}},\ and\ \bibinfo
  {author} {\bibfnamefont{L.}~\bibnamefont{Pesquera}},\ }%
  \bibfield{journal}{%
  \bibinfo {journal} {Zeitschr. f. Phys. B}\ }%
  \textbf{\bibinfo {volume} {52}},\ \bibinfo {pages} {335} (\bibinfo {year}
  {1983})%
  \bibAnnoteFile{NoStop}{Hernandez-Machado:ZFPBM:52:p335:1983}%
\bibitem{Kampen:BJoP:28:p90:1998}%
  \BibitemOpen
  \bibfield{author}{%
  \bibinfo {author} {\bibfnamefont{N.~G.}\ \bibnamefont{van Kampen}},\ }%
  \bibfield{journal}{%
  \bibinfo {journal} {Braz. J. Phys.}\ }%
  \textbf{\bibinfo {volume} {28}},\ \bibinfo {pages} {90} (\bibinfo {year}
  {1998})%
  \bibAnnoteFile{NoStop}{Kampen:BJoP:28:p90:1998}%
\bibitem{Kampen:StochProcPhysChem:Book:1981}%
  \BibitemOpen
  \bibfield{author}{%
  \bibinfo {author} {\bibfnamefont{N.~G.}\ \bibnamefont{van Kampen}},\ }%
  \emph{\bibinfo {title} {Stochastic Processes in Physics and Chemistry}}\
  (\bibinfo {publisher} {North-Holland},\ \bibinfo {address} {Amsterdam},\
  \bibinfo {year} {1981})%
  \bibAnnoteFile{NoStop}{Kampen:StochProcPhysChem:Book:1981}%
\bibitem{Novikov:SPJ:20:p1290:1965}%
  \BibitemOpen
  \bibfield{author}{%
  \bibinfo {author} {\bibfnamefont{E.~A.}\ \bibnamefont{Novikov}},\ }%
  \bibfield{journal}{%
  \bibinfo {journal} {Sov. Phys. JETP}\ }%
  \textbf{\bibinfo {volume} {20}},\ \bibinfo {pages} {1290} (\bibinfo {year}
  {1965})%
  \bibAnnoteFile{NoStop}{Novikov:SPJ:20:p1290:1965}%
\bibitem{DEKKER:PLA:90:p26:1982}%
  \BibitemOpen
  \bibfield{author}{%
  \bibinfo {author} {\bibfnamefont{H.}~\bibnamefont{Dekker}},\ }%
  \bibfield{journal}{%
  \bibinfo {journal} {Phys. Lett. A}\ }%
  \textbf{\bibinfo {volume} {90}},\ \bibinfo {pages} {26} (\bibinfo {year}
  {1982})%
  \bibAnnoteFile{NoStop}{DEKKER:PLA:90:p26:1982}%
\bibitem{Tserkovnyak:PRL:88:p117601:2002}%
  \BibitemOpen
  \bibfield{author}{%
  \bibinfo {author} {\bibfnamefont{Y.}~\bibnamefont{Tserkovnyak}}, \bibinfo
  {author} {\bibfnamefont{A.}~\bibnamefont{Brataas}},\ and\ \bibinfo {author}
  {\bibfnamefont{G.~E.~W.}\ \bibnamefont{Bauer}},\ }%
  \bibfield{journal}{%
  \bibinfo {journal} {Phys. Rev. Lett.}\ }%
  \textbf{\bibinfo {volume} {88}},\ \bibinfo {pages} {117601} (\bibinfo {year}
  {2002})%
  \bibAnnoteFile{NoStop}{Tserkovnyak:PRL:88:p117601:2002}%
\end{thebibliography}%
\bibliographystyle{apsrev4-1}

%
%
%
%
%
%
  
\end{document}